\documentclass[final,3p,times,twocolumn]{elsarticle}

\usepackage{hyperref}
\usepackage{amssymb}
\usepackage{orcidlink}
\DeclareUnicodeCharacter{2212}{-}
\usepackage{subcaption}
\usepackage{threeparttable}
\usepackage{etoolbox}
\appto\TPTnoteSettings{\footnotesize}
\usepackage{xcolor} 
\usepackage{multirow} 
\usepackage[export]{adjustbox}
\usepackage{comment}

\journal{Journal of Magnetism and Magnetic Materials}

\begin{document}

\begin{frontmatter}


\author{P.L.S. Cambalame\corref{cor1}\fnref{label1,label2} \orcidlink{0000-0002-6959-8931}}
\ead{phinifolo@fis.uc.pt}
 \author{B.J.C. Vieira \orcidlink{0000-0002-6536-9875}\fnref{label3}}
\author{J.C. Waerenborgh\orcidlink{0000-0001-6171-4099}\fnref{label3}}
\author{P.S.P. da Silva \orcidlink{0000-0002-6760-4517}\fnref{label1}}
\author{J.A. Paix\~{a}o \orcidlink{0000-0003-4634-7395}\corref{cor2}\fnref{label1}}
 \ead{jap@fis.uc.pt}

 \cortext[cor2]{}
 \affiliation[label1]{organization={CFisUC, Department of Physics, University of Coimbra},
             city={Coimbra},
             postcode={3004-516},
             country={Portugal}}
\affiliation[label2]{organization={Department of Physics, Eduardo Mondlane University}, 
           city={Maputo}, 
           postcode={0101-11}, 
           country={Mozambique}}
\affiliation[label3]{organization={Centro de Ci\^{e}ncias e Tecnologias Nucleares, DECN, Instituto Superior T\'{e}cnico, Universidade de Lisboa},
             city={Bobadela},
             postcode={2695-066},
             state={LRS},
             country={Portugal}}
\title{$\rm CrFe_2Ge_2$: Investigation of novel ferromagnetic material of $\rm Fe_{13}Ge_{8}$-type crystal structure} 

\begin{abstract}
We successfully synthesized a novel intermetallic compound $\rm CrFe_2Ge_2$ with the $\rm Fe_{13}Ge_{8}$-type crystal structure. A structural study is presented combining single-crystal X-ray diffraction and   M\"{o}ssbauer spectroscopy analysis, confirming the presence of two distinct Fe sublattices.  $\rm CrFe_2Ge_2$ exhibits a metallic ferromagnetic state with $T_C \approx \rm 200~K$.
This material does not follow the usual  $M^2 \propto H/M$ Arrott law, rather a modified Arrott law is obeyed in this material. The critical exponents determined from detailed analysis of modified Arrott plots were found to be $\beta = 0.392$, $\gamma = 1.309$ and $\delta = 4.26$ obtained from the critical isotherm at $ T_{\rm C} =\rm 200~K$. Self-consistency and reliability of the critical exponent analysis were verified by the Widom scaling law and scaling equations. Using the results from renormalization group calculation,  the critical behavior of $\rm CrFe_2Ge_2$ is akin to that of a $d=3, n=3$ ferromagnet in which the magnetic exhange distance is found to decay  as $J(r) \approx r^{-4.86}$  with long-range magnetic coupling. The evaluated Rhodes-Wohlfarth ratio of $\sim 3$ points to an itinerant ferromagnetic ground state. Low-temperature measurements of resistivity, $p(T)$, and specific heat, $C_P(T)$, reveal a pronounced contribution from electron-magnon scattering. 
\end{abstract}


\end{frontmatter}

\section{\label{sec:intro}Introduction}
The $\rm Fe_{13}Ge_{8}$-type crystal structure remains an underexplored benchmark for studying magnetic interactions. The crystal structure of $\rm Fe_{13}Ge_{8}$ is hexagonal (space group $P6_3/mmc$)~ \cite{MALAMAN1980155, Lidin:an0539}, featuring two trigonal pyramids of Ge atoms (at $6h$ Wyckoff position) facing opposite basal planes of $\rm Fe$ atoms. These $\rm Fe$ atoms comprise an octahedrally coordinated site ($2a$) and a partially occupied triangular sublattice of Fe atoms ($6h$) centered at the apex of each Ge-pyramid. Additionally, $\rm Fe$ atoms at ($6g$) position are octahedrally coordinated to $\rm Ge$ and square planar coordinated to the next-nearest neighbors.  This three-dimensional network of $\rm Fe$ atoms can be partially substituted by other transition-metal atoms,  enabling the system to host competing magnetic interactions. As such, it serves as an excellent platform for investigating the interplay between different magnetic exchange interactions and related effects such as spin canting, non-collinear magnetic ordering, competing competing ordering, and frustration \cite{OMoze_1994, ADELSON19651795, KHALANIYA2019118, khalaniya2021magnetic, Karna2021scfege, khalaniya2021magnetic, KITAGAWA2020121188, D5DT00654F}.

The previously synthesized $\rm Fe_{13}Ge_{8}$-type ferromagnetic compound $\rm Fe_3Ga_{1.7}As_{0.3}$~\cite{GREAVES1990315} was reported to exhibit a more ordered form of the $\rm B8_2$-type structure and a higher $T_{\rm C}$ than the initial compound $\rm Fe_3GaAs$\,~\cite{HARRIS1989103}. In subsequent work, Kitagawa obtained the same crystal structure in $\rm Fe_3Ga_{0.35}Ge_{1.65}$ which displays a ferromagnetic ground state~ \cite{Kitagawa2022}. The transition from a disordered $\rm Ni_2In$-type structure to a more ordered $\rm Fe_{13}Ge_{8}$ superstructure-like phase is not always achieved through doping or substitution of Ge/Fe. Structural and magnetic studies of $\rm Fe_{3.+{\it y}}{\it T}_{y}Ge_2$ ({\it T} = transition metals)\,\cite{kanematsu1963, Albertini1998} revealed that the substituting $\rm Fe$ by $\rm Mn$, $\rm Co$ and $\rm Ni$ reduces $T_{\rm C}$  and increases the axial anisotropy contribution with increasing $\it T$ content, though no structural phase transition was observed. Further {\it T}-Fe-Ge phase diagram studies revealed quite different structures and quantum phases. The $T\rm{Fe_2Ge_2}$ stochiometry with  $T \rm=Cu$ gives orthorhombic crystal structure with $\Delta$-Fe iron chains \cite{zavalij1987structure, may2016competing} and coupling of magnetic order and electrons similar to iron-based superconductors \cite{shanavas2015, BUDKO2018260}.  $\rm YFe_2Ge_2$ is tetragonal of $\rm ThCr_2Si_2$ structure \cite{Zou2014yfe2ge2}, as most $R\rm{Fe_2Ge_2}$ with {\it R} = alkaline earth, lanthanide or actinide \cite{BRAUN2019368,WELTER200335,EBIHARA1995219,AVILA200451},  yet $\rm YFe_2Ge_2$ is the only iron-germanide compound (FEG) displaying superconductivity, to our knowledge. Additionally, the implementation of $\rm Fe-Ge$ and ${\it T}\rm -Fe-Ge$ systems in thin-films and and nanowires results in higher magnetic ordering temperature \cite{Qin2011, Jiang2025}, exotic quantum states such as magnetic skyrmions and magnetic bubbles \cite{Yu2011-ph, Bei_Ding2020, Jiawang_Xu2024} establishing a great promise for potential applications in magnetic information storage and spintronics near room temperature \cite{Maat2008, Deng2018}. 

Most of the aforementioned compounds exhibit  a ferromagnetic ground state \cite{MALAMAN1980155, GREAVES1990315, Kitagawa2022, kanematsu1963, Albertini1998, BRAUN2019368}, reinforcing the correlation identified by \cite{KHALANIYA2019118, yasukochi1961magnetic} that $\rm Fe-Ge$ compounds with a more three-dimensional framework of $\rm Fe$ atoms tend to exhibit ferromagnetism. This is attributed to the dominance and nature of $\rm Fe-Fe$ direct exchange interactions over competing $\rm Fe-Ge-Fe$   antiferromagnetic interactions \cite{Kitagawa2022, Deiseroth2006, May2016fe3xgete2, Zhang2020fe5xget2, BRAUN2019368, Das1994kondo, etde_5753394, MAZET201379, VENTURINI199299, SCHOBINGERPAPAMANTELLOS199859}. In this context the investigation of the critical behavior of FEGs in the vicinity of the paramagnetic (PM) to ferromagnetic (FM) transition region can yield valuable insights into the magnetic exchange\,\cite{Liu2016-pz}, competing magnetic interactions\,\cite{Yu_Liu2020} and even disclose hidden phases\,\cite{Dara_2023}. The estimation of critical exponents can help sort out which of theoretical models: mean-field, Ising, XY or Heisenberg model best fit the magnetic systems as well as its spin dimensionality (1D, 2D or 3D). The universality classes, determined from critical exponents, are closely related to the crystal structure and spin correlation length, as observed in van der Waals (vdW) structures, as the magnetic critical behavior in quasi-2D vdW magnets CrSiTe$_3$ consistently suggest the universality class of the 2D Ising model,\cite{V_Carteaux_1995} while in $\rm CrGeTe_3$, with smaller vdW gap and stronger interlayer interaction, its critical behavior is closer to 3D tricritical mean-field class\,\cite{G_Lin2017}. In this context, the investigation of the critical behavior of  $\rm Fe_{13}Ge_{8}$-type compounds can yield important microscopic information about the underlying magnetic interactions. 

In this study, we investigate in detail single-crystal and polycristalline samples of $\rm CrFe_2Ge_2$ by single-crystal X-ray diffraction, M\"{o}ssbauer spectroscopy, electrical resistivity, specific heat, magnetic susceptibility, and isotherm magnetization measurements. The standard method of determining the Curie temperature from Arrott plots ($M^2 -H/M$) fails to give the correct  transition temperature. However, our results suggest that a modified Arrott plot based on the 3D-Heisenberg model is obeyed in this compound. Scaling analysis and renormalization group calculations confirms the consistency of the selected model.

\section{Experimental procedures}
Polycrystalline samples of $\rm CrFe_2Ge_2$ were synthesized using a solid state reaction method from a stoichiometric melt sealed quartz ampoule . The reactants were annealed at $\rm 610~^\circ C$ for 32 hours and then heated to $\rm 720~^\circ C$ and kept at this temperature for 18 hours, and finally heated to $\rm 1050~^\circ C$ and kept at this temperature for 18 hours, followed by cooling at a rate of $\rm 0.1~^\circ C/min$ in the $\rm 1050-850~^\circ C$ range. Other samples were obtained by a modified route, details in the \textit{Supplementary Information}.
Structural characterization was performed using single-crystal X-ray diffractometry on a Bruker ApexII diffratometer, with a kappa goniometer, equipped with a 4K CCD detector and a Mo X-ray tube with graphite monochromator. The powder diffraction data were collected on a Bruker D8 Advance X-ray powder diffractometer using Ni-filtered Cu K$\alpha$ radiation and a silicon-drift LynxEye one-dimensional detector. 

The crystal structure determined from the single-crystal data was refined using SHELXL~\cite{sheldrick2015crystal}. Structure refinement from powder diffraction data were refined using Profex 4 ~\cite{Doebelin:kc5013}. Sample composition was analyzed using energy dispersive X-ray spectrometer (EDXS). Magnetic (VSM) measurements were performed on a Quantum Design Dynacool Physical Property Measurement System (PPMS) equipped with a 9 T superconducting magnet. Resistivity and specific heat measurements were also performed in this equipment, using the Quantum Design resistivity bridge and specific heat (relaxation method) plug-in options.

M\"{o}ssbauer spectra were collected at room temperature and 80\,K in transmission mode using a conventional constant-acceleration spectrometer and a 50 mCi $^{57}$Co source in a Rh matrix. The velocity scale was calibrated using $\alpha$-Fe foil. Isomer shifts (IS) are given relative to this standard at room temperature. The absorber was obtained by gently packing the sample in a perspex holder. Absorber thickness was calculated on the basis of the corresponding electronic mass-absorption coefficients for the 14.4 keV radiation, according to Long et al. \cite{long1983ideal}. The low-temperature measurement was performed in a bath cryostat with the sample immersed in He exchange gas. The spectrum was fitted to Lorentzian lines using a non-linear least-squares method. The relative areas and line widths of both peaks in each doublet and of peaks 1-6, 2-5 and 3-4 in each magnetic sextet were constrained to remain equal during the refinement procedure, as expected for samples with no texture effects.

\section{Results}
\subsection{Single crystal and powder X-ray diffraction}
\begin{figure*}[htpb]
	\captionsetup{font=normalsize}
	\centering
	\begin{subfigure}{.4\linewidth}
		\centering
		\includegraphics[scale=0.25,trim={95 5cm 50 250}]{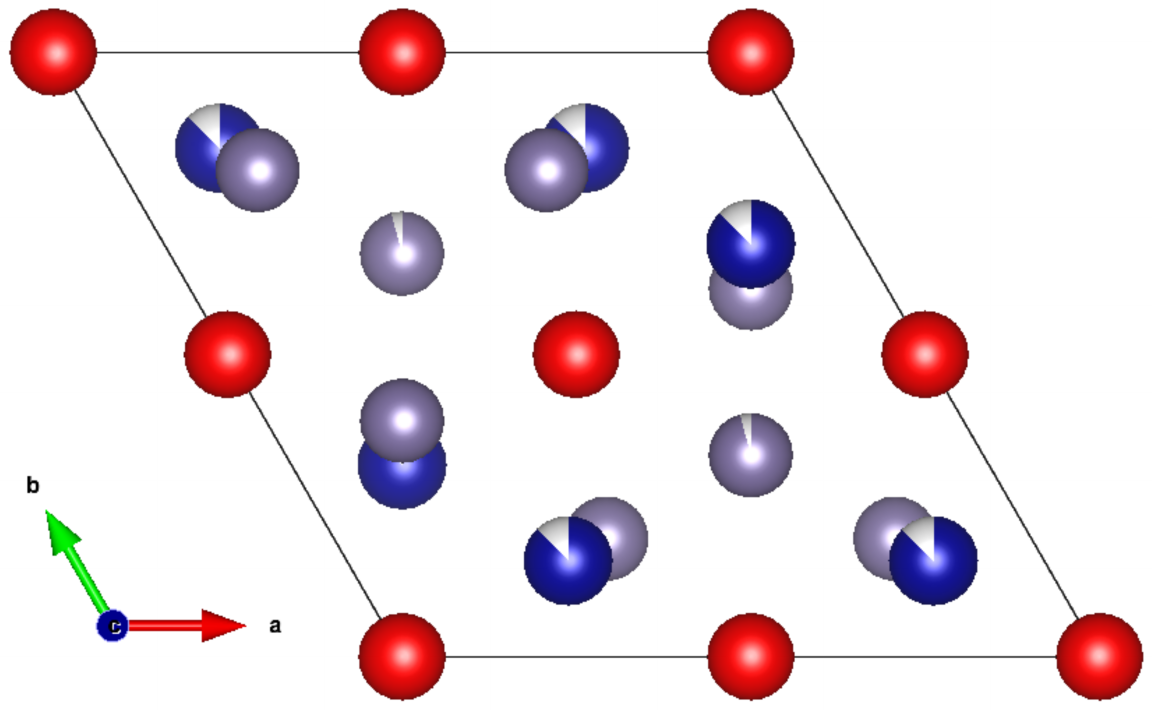}
		\caption{}
		\label{fig:fig_dos}
	\end{subfigure}%
	\begin{subfigure}{.4\linewidth}
		\centering
		\includegraphics[scale=0.25,trim={75 5cm 40 200}]{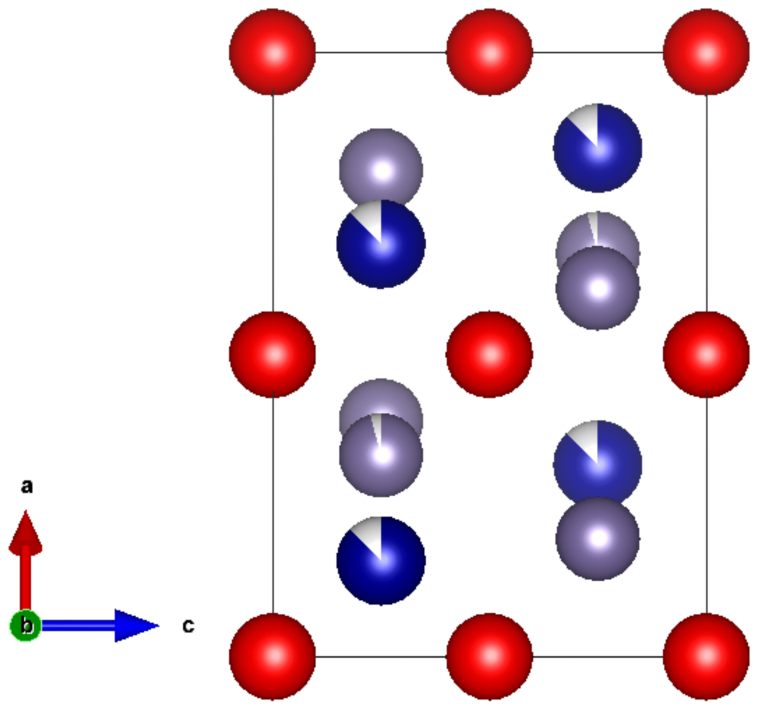}
		\caption{}
	\end{subfigure}%
	\begin{subfigure}{.22\linewidth}
		\includegraphics[width = 0.75\linewidth, inner]{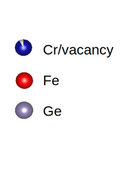}
	\end{subfigure}%
	\caption{Crystal structure of  CrFe$_{2}$Ge$_{2}$ projected along the \textit{c} axis (a) and \textit{a} axis (b). }
	\label{fig:scxrd}
\end{figure*}
The new material was investigated by both single-crystal and powder X-ray diffraction (PXRD). The compound $\rm CrFe_2Ge_2$ crystallizes in the hexagonal $P6_3/mmc$ space group, with a $\rm Fe_{13}Ge_8$-type~\cite{MALAMAN1980155} type of structure. It is a superstructure of the closely related $\rm Fe_{1.67}Ge$ compound~\cite{laves1942einige,yasukochi1961magnetic}, with a doubled $a$ lattice parameter. In $\rm CrFe_2Ge_2$, all Fe atoms that (partially) occupy the $6h$ positions of the parent $\rm Fe_{13}Ge_8$ structure are fully replaced by Cr atoms, whereas the Ge atoms keep their positions. The superstructure consists  of four  alternating Cr-rich and Cr-poor  subunits, with approximate compositions $\rm Cr_{\it x}FeGe$ ($x = 1$ and 0.5). 
The unit cell contains two alternating trigonal pyramids of Ge along the $<110>$ directions, as seen in a projection in the $ab$ plane depicted in Fig. \ref{fig:scxrd}. The atoms were found to be distributed among five distinct crystallographic sites. A summary of the single-crystal data collection and structure refinement is provided in Table \ref{tab:table_sxrd} and \ref{tab:tablesc} and the crystal structure is displayed in  Fig. \ref{fig:scxrd}.  
Assuming that the $2a$ sites are fully occupied by Fe atoms, it was found that the $6g$ positions have close to full occupancy of Fe atoms and the $6h$ positions have an occupancy of only $\rm 88\%$ of Cr atoms. 

The Fe-atoms at the $6g$ site (Fe1)  and the $2a$ sites (Fe2) are both octahedrally  coordinated to Ge and Cr, respectively. The octahedra centered at 6\textit{g} has a stronger trigonal distortion, with Ge-Ge distances varying from $3.174\rm\textup{~\AA}$~to~$3.704\rm\textup{~\AA}$ and Fe-Ge distances, from 2.501 to $2.649\rm\textup{~\AA}$. The octahedra centered at Fe2, has all Cr atoms equidistant from Fe2, with Fe-Cr distances of $2.555\rm\textup{~\AA}$. Fe1 is, additionally, square-planar coordinated to Cr,  with a distance of $2.697\rm\textup{~\AA}$. The presence of two Fe-sublattices is further supported by M\"{o}ssbauer spectroscopy, as shown below. 
The $\rm Cr-Ge$ distances are around $\rm 2.560\textup{~\AA}$. It should be noted that the chemical composition as obtained by refinement of the crystallographic occupational parameters based on single crystal X-ray data gives $\rm Cr_{1.34} Fe_{2 }Ge_{1.98}$ instead of the nominal $\rm CrFe_2Ge_2$. For the nominal composition, all atomic positions except $6h$ are fully occupied and the position $6h$ should be fractionally occupied with a value of 0.67. The observed excess of electron density at the $6h$ site could be due to either an excess of Cr, or partial occupation of this site by the other atomic species.
One has to bear in mind, however, that refinement of occupancies from X-ray diffraction data of elements with large and similar number of electrons, such as the case of Cr and Fe, is not very reliable, an additional problem being a high correlation between site occupation and atomic displacement parameters (also known as temperature factors).  

Therefore, to further verify the stoichiometry and homogeneity of the sample, several small pieces were extracted from the bulk and analyzed using EDS. Elemental mapping confirmed good homogeneity and a composition close to $\rm Cr_{0.97}Fe_{2.06}Ge_{1.97}$. 

\begin{table}    	
	\captionsetup{font=small}
\centering
\caption{Single crystal X-ray diffraction data for  $\rm CrFe_{2}Ge_{2}$.}
	\begin{tabular}[t]{ll} \hline\hline
Chemical formula            & $\rm CrFe_{2}Ge_{2}$    \\
Molecular weight $\rm (g/mol)$ & 308.946 \\
Space group              &  $P6_3/mmc$ (194)      \\
$a$ (\AA)                &  8.0445(2)                \\
$b$ (\AA)                &  8.0445(2)                \\
$c$ (\AA)                &  5.00290(10)                \\
$V$(\AA$^3$)             &  280.38231             \\
Temperature (K) & 300 \\
$\rho_{\rm calc} ~(\rm g cm^{-3})$           &          7.617     \\
Absorption coefficient $\mu ~ (\rm mm^{-1})$ &   35.366  \\
Absorption corrections $T_{min}$, $T_{max}$  &  0.3361, 0.7474    \\
Data collection range $(^{\circ})$           &     $2.924 \le \theta \le  37.812$\\
$h$ range                                    &   $-13\le h \le  13$\\
$k$ range                                    &  $-13 \le k \le 13$\\
$l$ range                                    &  $-8 \le l \le 8$\\
Reflections collected                        & 56658 \\
Independent reflections                      & 320 \\
Parameters refined                           & 21 \\
$R_{\rm int}$                                           & 0.0942\\
$R_{1}(F)$ for $I > 2\sigma\, ^{a}$                             & 0.0392 \\
$wR_{2}(F_o^2)^b$ for all data                                       &  0.0733\\
Goodness-of-fit on $F^2$   & 1.091\\
CSD $\#$   & 2466680\\
\hline
\footnotesize $^{a} R_{1} = \sum ||F_{o}|-|F_{c}||/\sum|F_{o}||$. \\
 \multicolumn{2}{l}{\footnotesize $^{b} wR_{2} = [ w(F_{o}^{2}-F_{c}^{2})^2/\sum w(F_{o}^{2})^2]^{1/2}$,} \\
\multicolumn{2}{l}{\footnotesize $w=1/[\sigma^2(F_{o}^{2})+(A\cdot  P)^2+B\cdot P]$,} \\
\multicolumn{2}{l}{\footnotesize $P=[2F_{c}^2 + Max(F_{o}^2,0)/3]$ where} \\  \multicolumn{2}{l}{\footnotesize $A = 0.0487$ and $B=10.442$}  \\
\label{tab:table_sxrd}
	\end{tabular}
	\end{table}

\begin{figure}[!htpb]
	\captionsetup{font=small}
	\centering
	\includegraphics[width = \linewidth]{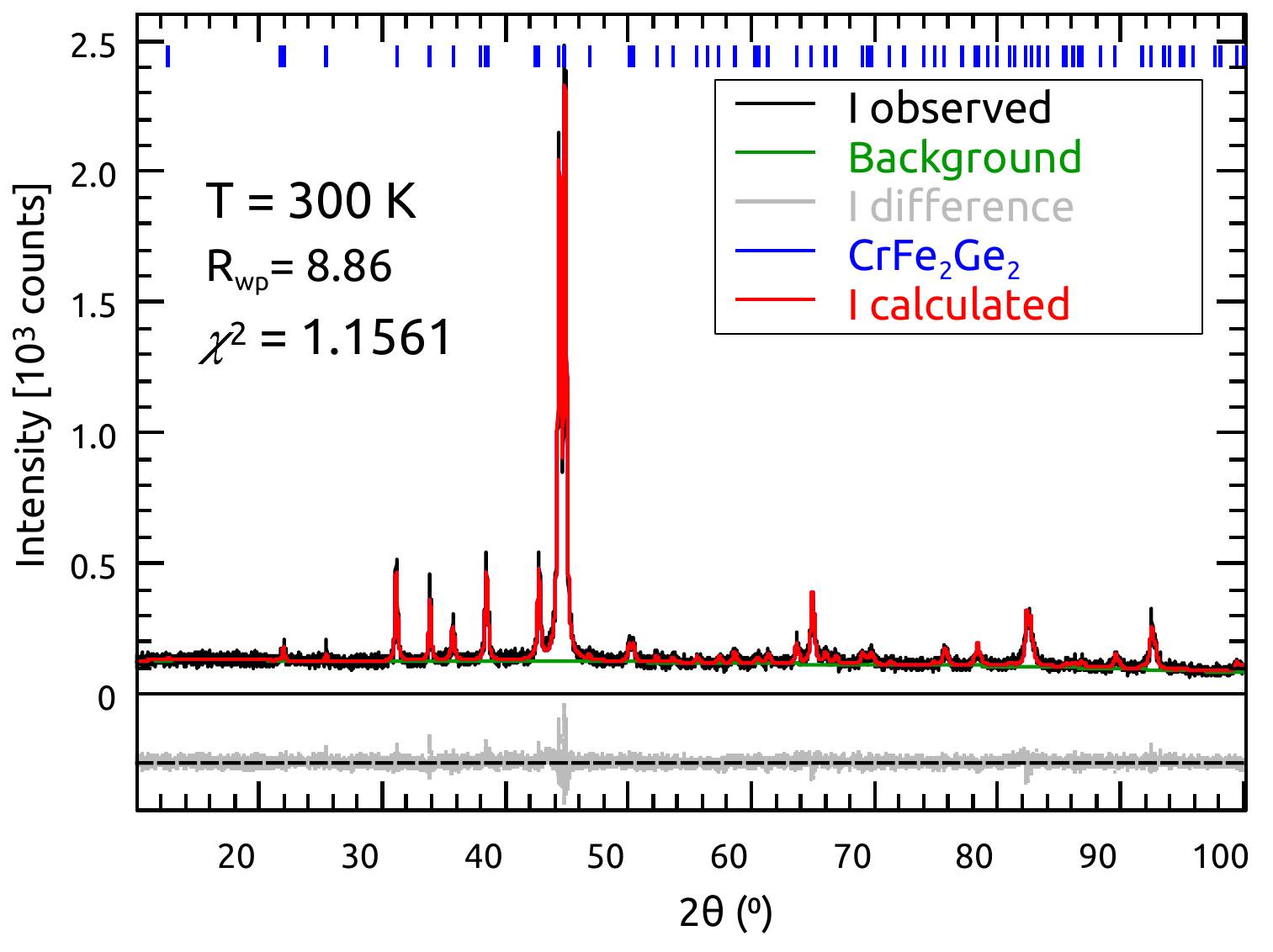}
	\caption{PXRD $\rm CrFe_2Ge_2$ Rietveld refinement. Blue tick marks indicate X-ray diffraction peaks of $\rm CrFe_2Ge_2$.} 
	\label{fig:pxrd}
\end{figure}

\begin{table*}  
		\captionsetup{font=small}  
\centering
\caption{Atomic coordinates, equivalent displacement parameters and occupancies of $\rm CrFe_{2}Ge_{2}$ at 300 K. The number in square brackets indicates the number of of bonds of symmetry equivalents}
	\begin{tabular}[t]{cccccccc} \hline
  & \textit{x} & \textit{y} &\textit{z}  & $U_{\rm eq}\rm(\textup{\AA}^2)$ & Occupancy & site  \\
		\hline
            Ge1     &    0.39032(18)     &  0.19516(9)        & 0.25000       & 0.0152(3) &1.000    & \textit{6h} \\
            
            Ge2     &  0.33333       & 0.66667         & 0.25000       
            &  0.0208(6) & 0.973(12)     & \textit{2c}   \\
            
            Fe1     &    0.50000      & 0.00000        & 0.00000        & 0.0079(3) & 0.985(8)     & \textit{6g}   \\
            Fe2     &   0.00000       & 0.00000        & 0.00000        & 0.0236(6) & 1.000     & \textit{2a}   \\ 
            Cr1     &  0.15874(15)      & 0.3175(3)      & 0.25000       &  0.0170(5) & 0.880(10)   & \textit{6h}   \\
            
\hline
  Bond[Multiplicity]            & Fe1- Fe1[2]        & Fe1-Cr1[4]    & Ge2-Fe1 [2]     &Ge1-Fe1[4]  &  Fe2-Cr1 [6]\\
Bond lengths (\AA)              & 2.50145(5)            &   2.6882(13)        &  2.63764(5)        &    2.4897(5)     &  2.541(2) \\
 \hline
\label{tab:tablesc}
	\end{tabular}
	\end{table*}

The PXRD study of the sample revealed no detectable secondary phases. The results of the Rietveld refinement, using the structural model obtained from single-crystal X-ray diffraction as a starting point, are shown in Fig. \ref{fig:pxrd} (a). The observed and calculated PXRD patterns exhibit excellent agreement, with quality factors of $R_{wp}=8.96\%$ and $\chi= 1.18$.  A satisfactory Rietveld refinement was achieved by considering a degree of intersite disorder at the  $6h$ Wyckoff position, involving $\rm Ge/Fe$ and $\rm Cr/Ge$ substitutions at the $\rm Ge {\it (6h)}$ and $\rm Cr {\it (6h)}$ sites, respectively.

\subsection{M\"{o}ssbauer Measurements} 
\begin{figure}[!htpb]
	\captionsetup{font=small}
	\centering    \includegraphics[width=0.5\textwidth]{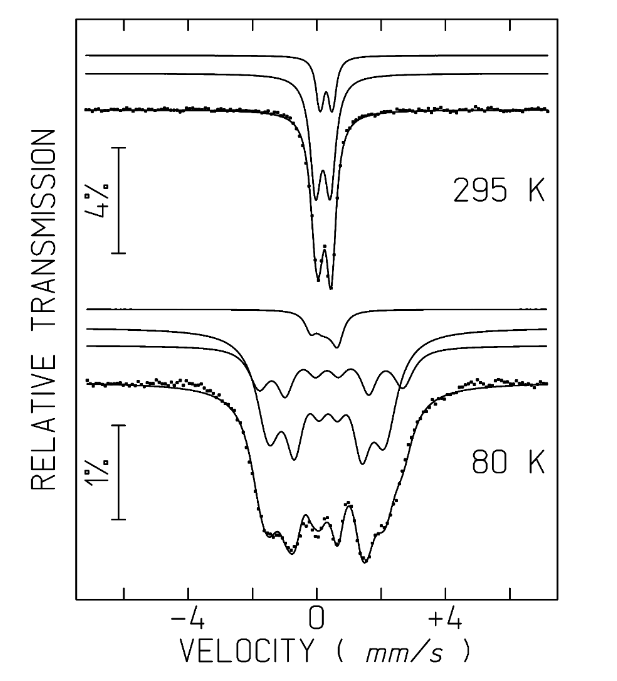}
	\caption{M\"{o}ssbauer 295 K and 80 K spectra of CrFe$_{2}$Ge$_{2}$  system. The calculated lines over the experimental points are the sum of two doublets	and three sextets, respectively (see Table \ref{tab:mossb}), shown slightly shifted for clarity. }
	\label{fig:mossb}
\end{figure}
\begin{table*}
	\centering
	\captionsetup{font=small}
	\caption{M\"{o}ssbauer parameters characterization. Isomer shift (IS), quadrupole splitting (QS), hyperfine field (B$_{hf}$), line-width ($\Gamma$) and relative area (I).}
	\begin{threeparttable}[t]
		\centering
		\begin{tabular}[t]{|c|c|c|c|c|c|c|}
			\hline
			Sample   &  IS \tnote{1} (mm/s)& $\epsilon$, QS\tnote{2} \ (mm/s) & B$_{hf}$(T) & $\Gamma$ (mm/s) &  I  & Fe Position \\
			\hline
			\multirow{2}{*}{295 K}  & 0.36 & 0.30 & - & 0.26 &25 \% & \textit{2a} \\
			& 0.32 & 0.50  & - & 0.40 & 75 \% & \textit{6g}\\
			\hline
			\multirow{3}{*}{80 K}                 & 0.49 & 0.13 &13.9 & 0.63 & 25 \% & \textit{2a}  \\
			& 0.44 & -0.04 & 11.2 & 0.70 &69 \% & \textit{6g}\\
			& 0.42 & -0.15  & 2.6 & 0.41 & 6 \% & \textit{6g} \\
			\hline
		\end{tabular}
		\begin{tablenotes}
			\item[1] IS - isomer shift relative to metallic $\alpha-$Fe at 295 K;
			\item[2] QS -quadrupole splitting and $(2\epsilon) = (e^{2}V_{zz}Q/4)(3cos^{2}\theta-1)$ quadrupole shift estimated for quadrupole doublets and magnetic sextets, respectively. 
			\item[] Estimated errors $\leq 0.02 \rm mm/s$ for IS, QS, $\epsilon$, $<$ 0.2 T for B$_{hf}$  and $< 2\%$ for I.
		\end{tablenotes}
	\end{threeparttable}%
	\label{tab:addlabel}%
	\label{tab:mossb}
\end{table*}
In the M\"{o}ssbauer spectrum of  $\rm CrFe_2Ge_2$  obtained at room temperature, two absorption peaks are observed (Fig. \ref{fig:mossb}). According to crystallographic data  $\rm CrFe_2Ge_2$ crystallizes in the $P6_3/mmc$ space group ($nr$ 194) and Fe fully occupies two equipositions, \textit{6g} and \textit{2a}, with symmetry lower than cubic. The  $\rm CrFe_2Ge_2$ was therefore refined considering two quadrupole doublets. A good fit is obtained if the intensity ratios of both peaks is 3:1, in agreement with the multiplicity of the equipositions. Fe atoms on \textit{6g} sites have therefore a slightly lower isomer shift, IS, and a higher quadrupole splitting, QS, than Fe on \textit{2a} sites (Table \ref{tab:mossb}). 

The spectrum obtained at 80\,K, below the magnetic transition temperature, reveals the presence of strong magnetic correlations. The spectrum is however very complex, apparently resulting from the overlapping of several low-split sextets. Since these sextets are not resolved the analysis of the spectrum is not unique. On the other hand, at least three magnetic sextets are necessary to obtain a reasonable fit. Therefore an analysis consistent with room temperature results was performed assuming three magnetic sextets. The Fe on \textit{2a} sites with the highest number of Cr nearest neighbors have the highest  hyperfine field ($B_{hf}$) and consequently the highest magnetic moments. 
Fe atoms with different $B_{hf}$ on \textit{6g} sites may be related to the presence of both Cr  atoms and vacancies on the \textit{6h} nearest neighbour sites leading to different nearest neighbour environments of Fe atoms on \textit{6g} sites.

\subsection{Magnetic Measurements}
\begin{figure*}[htpb]
	\captionsetup{font=small}
	\centering
	\begin{subfigure}{.5\linewidth}
		\centering
		\includegraphics[width = .98\linewidth]{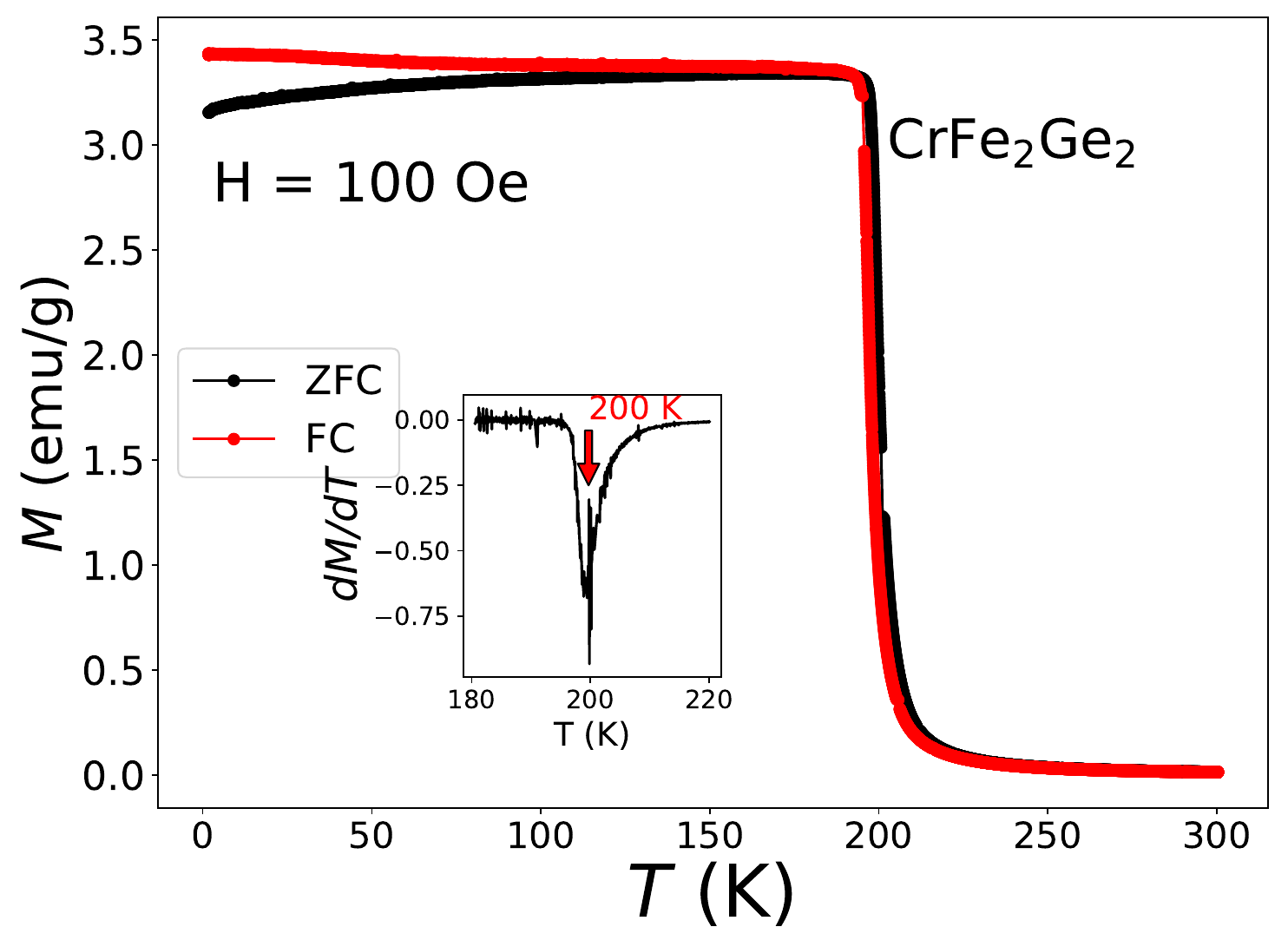}
		\caption{}
	\end{subfigure}%
	\vspace{0.41em}
	\begin{subfigure}{.5\linewidth}
		\centering
		\includegraphics[width =0.98\linewidth]{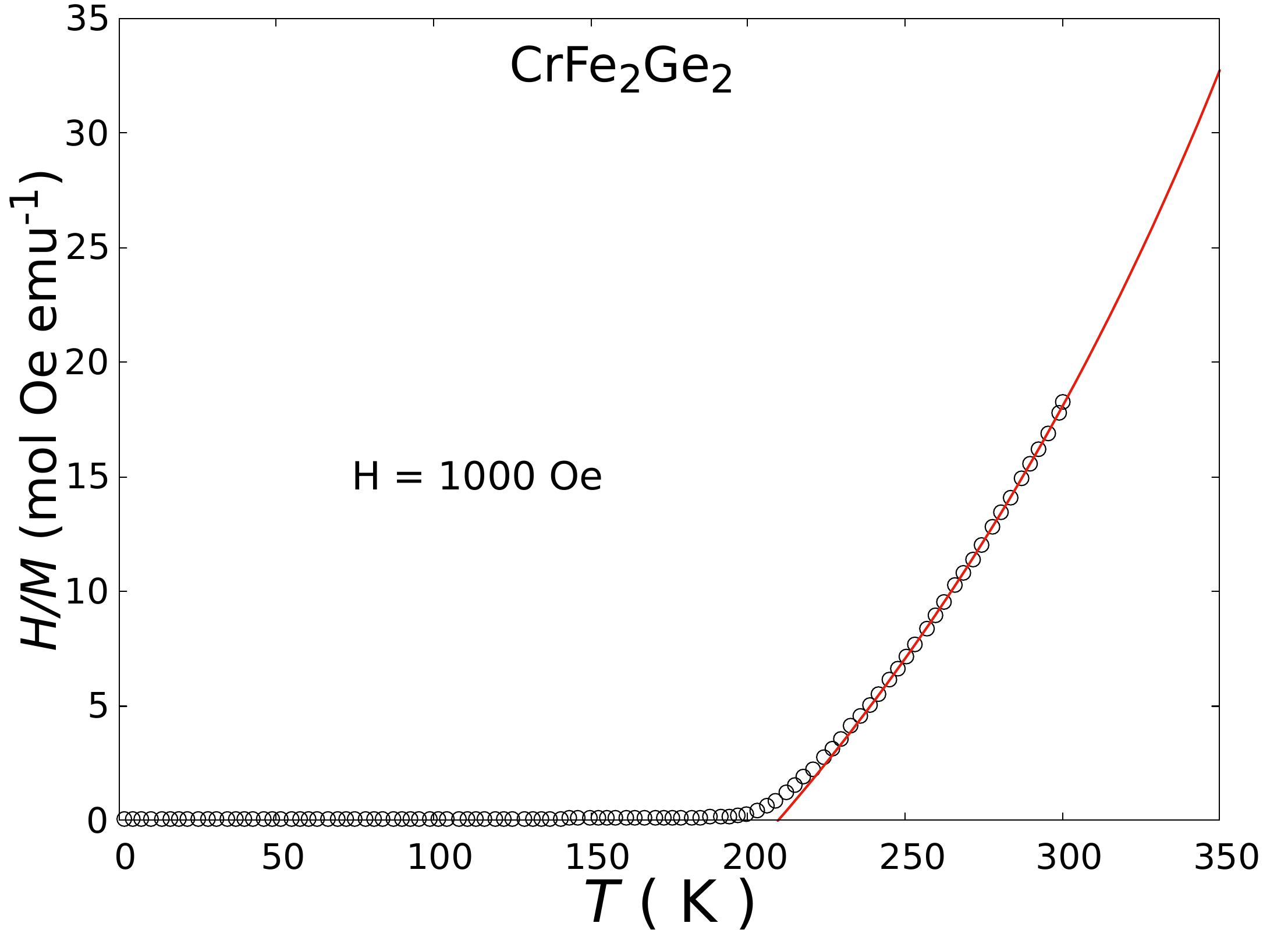}
		\caption{}
	\end{subfigure}%
	
	\hspace{1em}
	\begin{subfigure}{.5\linewidth}
		\centering
		\includegraphics[width =0.98 \linewidth]{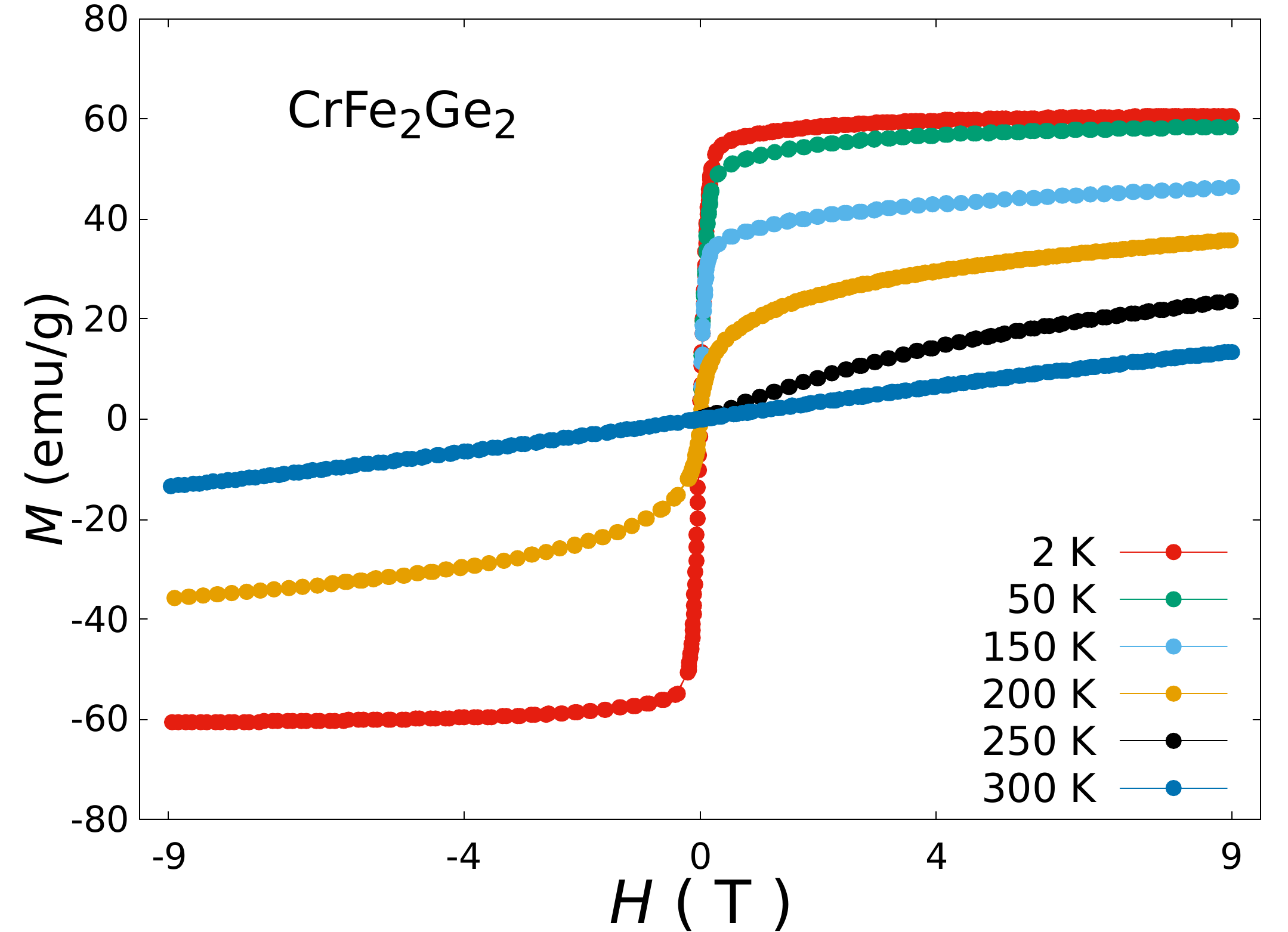}
		\caption{}
	\end{subfigure}%
	\vspace{0.41em}
	\begin{subfigure}{.5\linewidth}
		\centering
		\includegraphics[width = 0.98\linewidth]{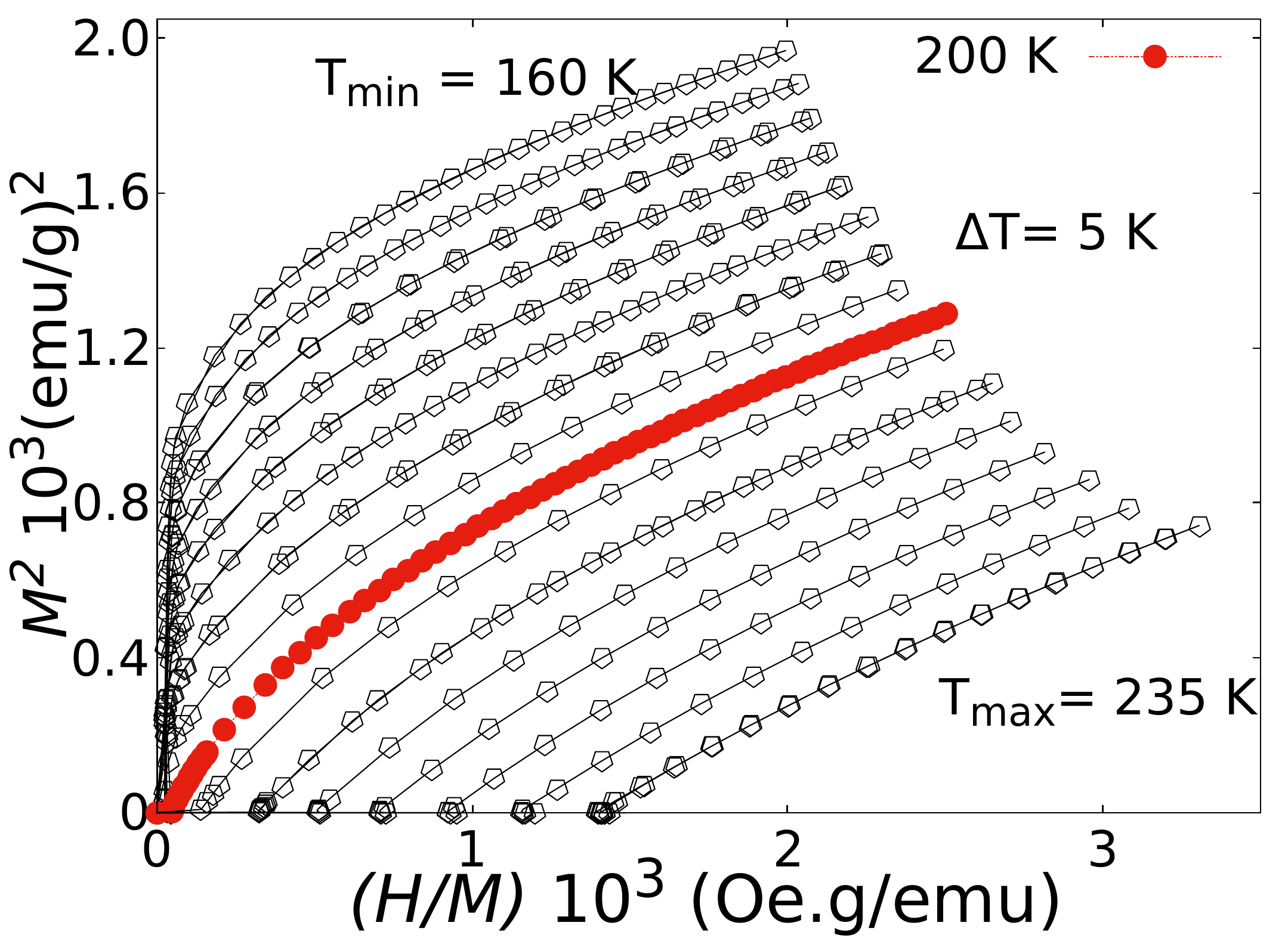}
		\caption{}
	\end{subfigure}%
	\caption{(a) Magnetization as function of temperature \textit{M(T)} of $\rm CrFe_2Ge_2$ and $dM/dT$ in the inset. (b) Temperature dependence of inverse magnetic susceptibility for $\rm CrFe_2Ge_2$.The solid lines indicate the ﬁt with the modified Curie–Weiss law. (c) Isothermal magnetization curves \textit{M(H)} at various temperatures at magnetic field \textit{H} of up to 9 T. (d) $M^2$ vs. $H/M$ plots (Arrott plot) for various temperature for $\rm CrFe_2Ge_2$.\label{fig:mag_data}}
\end{figure*}
\begin{figure}[htpb]
	\captionsetup{font=small}
	\centering    \includegraphics[width=0.5\textwidth]{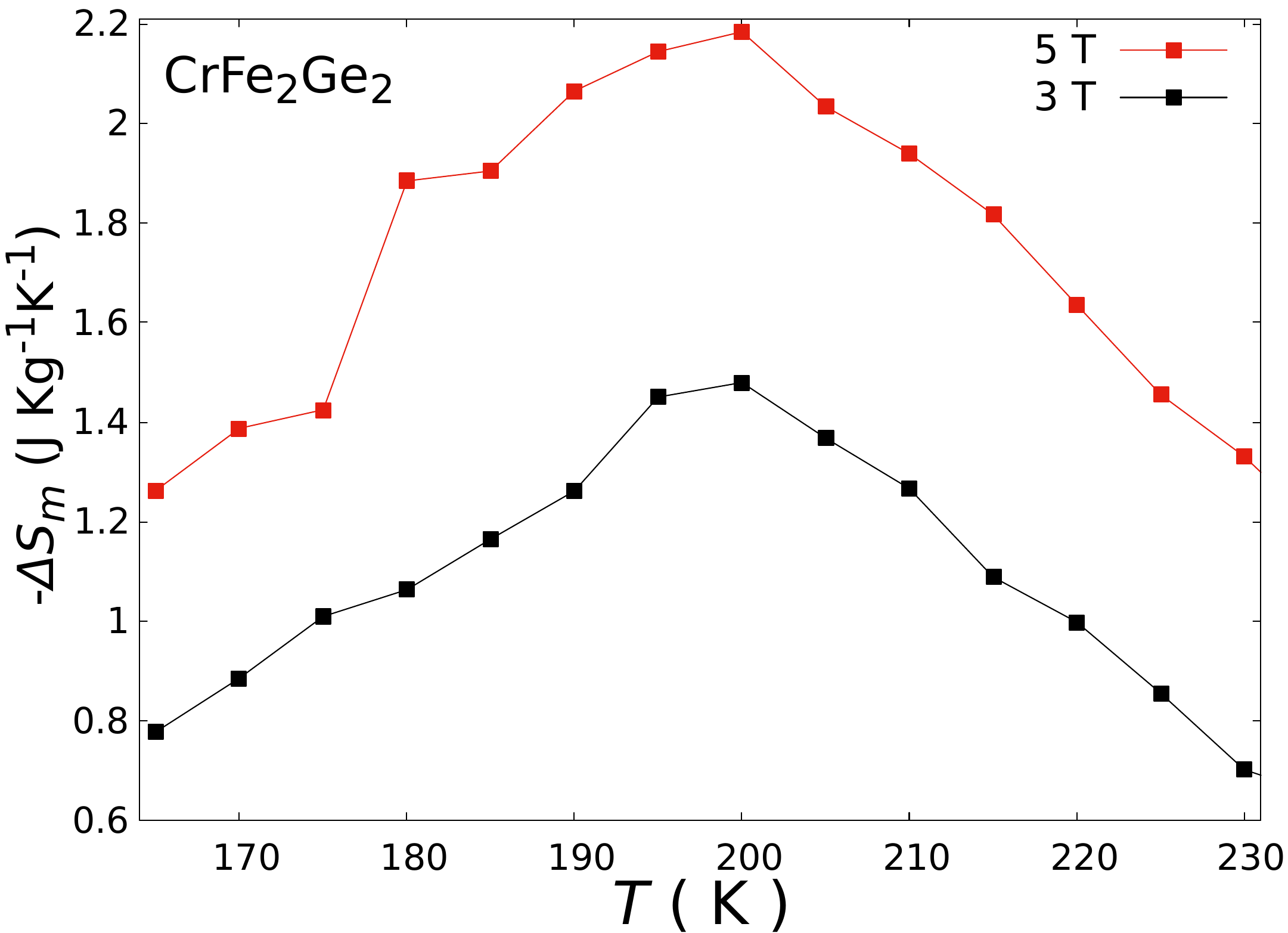}\caption{Temperature-dependent entropy changes for CrFe$_{2}$Ge$_{2}$}
	\label{fig:entrop}
\end{figure}
We carried out zero-field (ZFC) and field cooled (FC) magnetic measurements from $\rm 300~K$ down to $\rm 1.8~K$ in an applied field ranging from 50 to $\rm 1000~Oe$ in a polycrystalline sample.  The compound   $\rm CrFe_2Ge_2$  was found to be ferromagnetic with a transition temperature around $\rm 200~K$, estimated from $dM/dT$, Fig. \ref{fig:mag_data} (a). 

Figure \ref{fig:mag_data} (b) displays the inverse magnetic susceptibility. The linear range $[\mathrm{250 K - 300 K}]$ of the paramagnetic part of the inverse susceptibility $\chi ^{-1}$ was fitted to the modified Curie-Weiss law

\begin{equation}
    \chi = \frac{C}{T - \theta_p} + \chi_{0}
\end{equation}
where $C$ is the Curie-Weiss constant, $\theta_p$ is the Weiss constant and $\chi_{0}$ is the temperature independent susceptibility. The values of the parameters obtained from the fitting are $\theta_p =\rm  209.3(3)~K$ and $C = \rm 6.343(6)~emu~K/mol$ per formula unit ($\rm CrFe_2Ge_2$), $\chi_0=\rm -0.0144(5)~cm^3/mol$. The positive value of $\theta_p$ accounts for the ferromagnetic ordering of the spins. The effective magnetic moment per $3d$ atoms, calculated from the Curie-Weiss constant is $\rm 4.11 ~ \mu_B$ per $3d$ atoms. 
Using the relation $\mu _{eff}^2 = p_C(p_C + 2)$, the number of magnetic carriers per magnetic ion is $p_C =$ 3.23.
Furthermore, we inferred the itineracy of this system from the Rhodes-Wohlfarth ratio (RWR)\,\cite{rhodes1963effective} by comparing the saturation magnetic moment per $3d$ atoms calculated at $\rm 2~K$ by the linear extrapolation of $M^2$ to $H/M = 0$ and obtained $M_S =\rm 1.07~ \mu_{\rm B}$ per \textit{3d} atoms. The ratio $p_C/p_S$ gives 3, from the presented framework $\rm CrFe_2Ge_2$ is an itinerant ferromagnet. Presuming the applicability of Landau mean-field theory to this material, the critical temperature $T_{\rm C}$ of itinerant ferromagnet could be determined from the Arrott plot ($M^2$ vs. $H/M$) \cite{Arrot_1957}. The Arrott plot would show parallel straight lines, and one line should pass the origin as $T$ approaches $T_{\rm C}$. Figure \ref{fig:mag_data}(d) shows the $M^2$ vs. $H/M$ curves at various temperatures. $M^2$ does not show a linear relation but features a curvature at all temperatures under 200 K and a more straight behavior as the temperature get higher.  Similar behavior has been observed in materials such as $\rm Fe_3GeTe_2$ \cite{chen2013magnetic}, $\rm CrGeTe_3$ \cite{G_Lin2017}, $\rm LaCoAsO$ \cite{Ohta2009} and $\rm Fe_{\it x}Co_{1-\it x}Si$ \cite{Shimizu1990}.  Nevertheless, it's still possible to identify the nature of magnetic transition from the  Banerjee’s criterion \cite{BANERJEE196416}, as the positive slope corresponds to a second-order transition and the negative slope would correspond to first-order. The shape of the isotherms close to and into the FM region clearly indicates that the PM-FM transition in $\rm CrFe_2Ge_2$ is of second-order.

The field-dependent magnetization isotherms at $\rm 300~K$, 250 K, 200 K, 150 K, 50 K and 2 K confirm  the FM behavior of  $\rm CrFe_2Ge_2$ (Fig. \ref{fig:mag_data} (c)). At 2 K, $\rm CrFe_2Ge_2$ behaves as a soft ferromagnet with coercive field of only $\sim\rm  20~Oe$ and the magnetization shows saturation behavior above 35 kOe. With increasing temperature, the saturation magnetization $M_s$ decreases gradually and becomes proportional to $H$ around 250 K.  
We calculated the magnetic entropy change $(\Delta S_{mag})$ for  $\rm CrFe_2Ge_2$ by approximating the integral Maxwell equation with a sum:

\begin{equation}
    \Delta S_{mag}(T,\Delta H) =\sum _{0}^{H_{\rm max}} \frac{M_{i}-M_{i-1}}{T_{i} -T_{i-1}} \Delta H
\end{equation}
where $H_{\rm max}$ is the maximum external field. Experimental $M(H)$ curves from 165 to 230 K with increment of  $\Delta T = T_{i} -T_{i-1}$ = 5 K  were used for these calculations, resulting in negative values of  $\Delta S_{mag}$ (Fig. \ref{fig:entrop}). The $-\Delta S_{mag}$ value reaches its maximum around $T_{\rm C}$. For $H_{\rm max} =\rm 5~T$ and $H_{\rm max} = \rm 3~T$ the $-\Delta S_{mag}$  is less than $\rm 2.2 ~J/(Kg~ K)$ and  $\rm 1.4 ~J/(Kg~ K)$, respectively, which would be comparable to other isostructural compounds such as $\rm Cu_{0.6}Mn_{2.4}Ge_{2}$\,\cite{TENER2021167827} and $\rm Fe_{3}Ga_{0.35}Ge_{1.65}$\,\cite{Kitagawa2022}.

\subsection{Resistivity and heat capacity measurements}
\begin{figure*}[htpb]
	\captionsetup{font=small}
	\centering
	\begin{subfigure}{.5\linewidth}
		\centering
		\includegraphics[width = .974\linewidth]{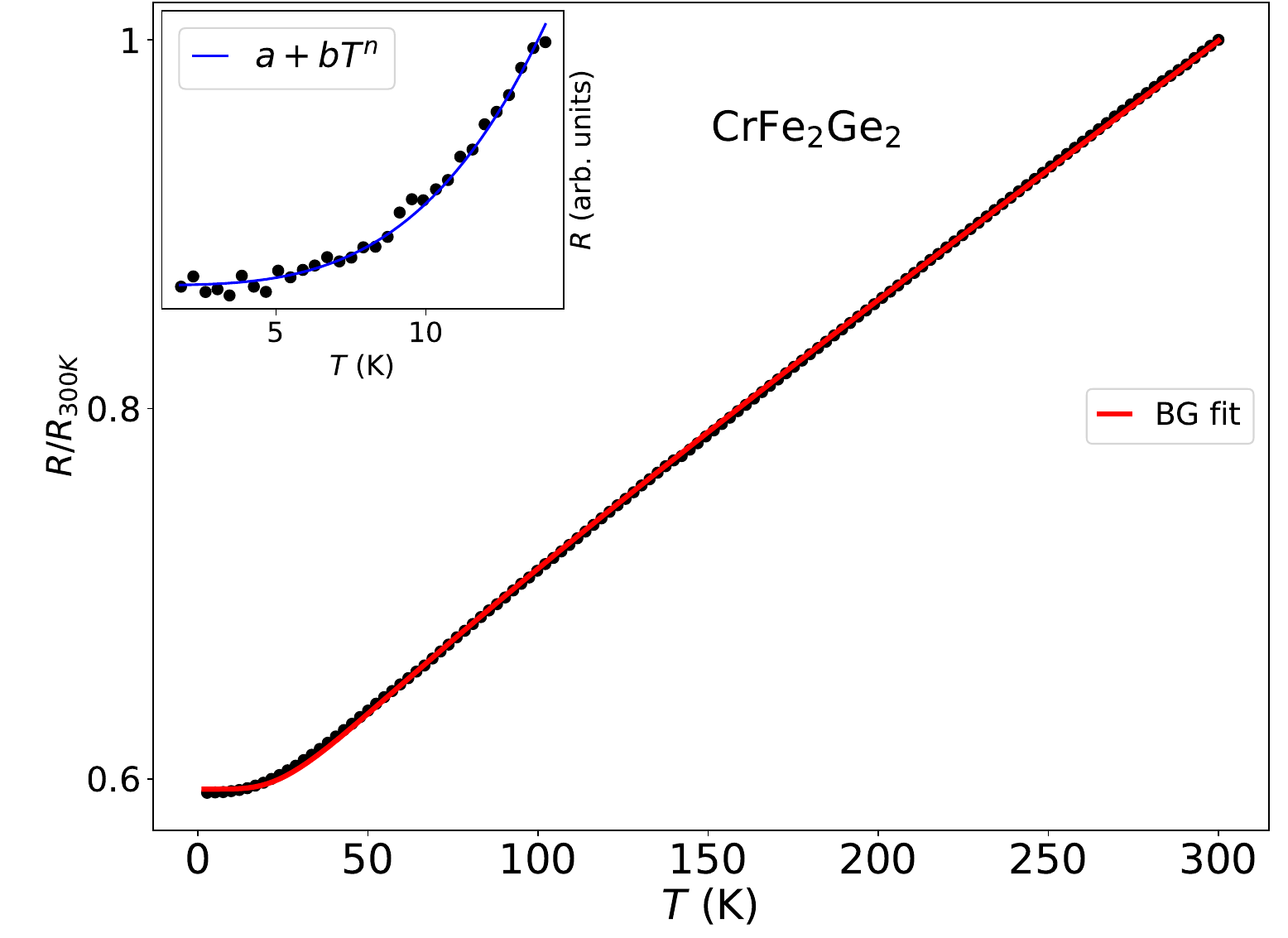}
		\caption{}
	\end{subfigure}%
	\begin{subfigure}{.5\linewidth}
		\centering
		\includegraphics[width =0.975\linewidth]{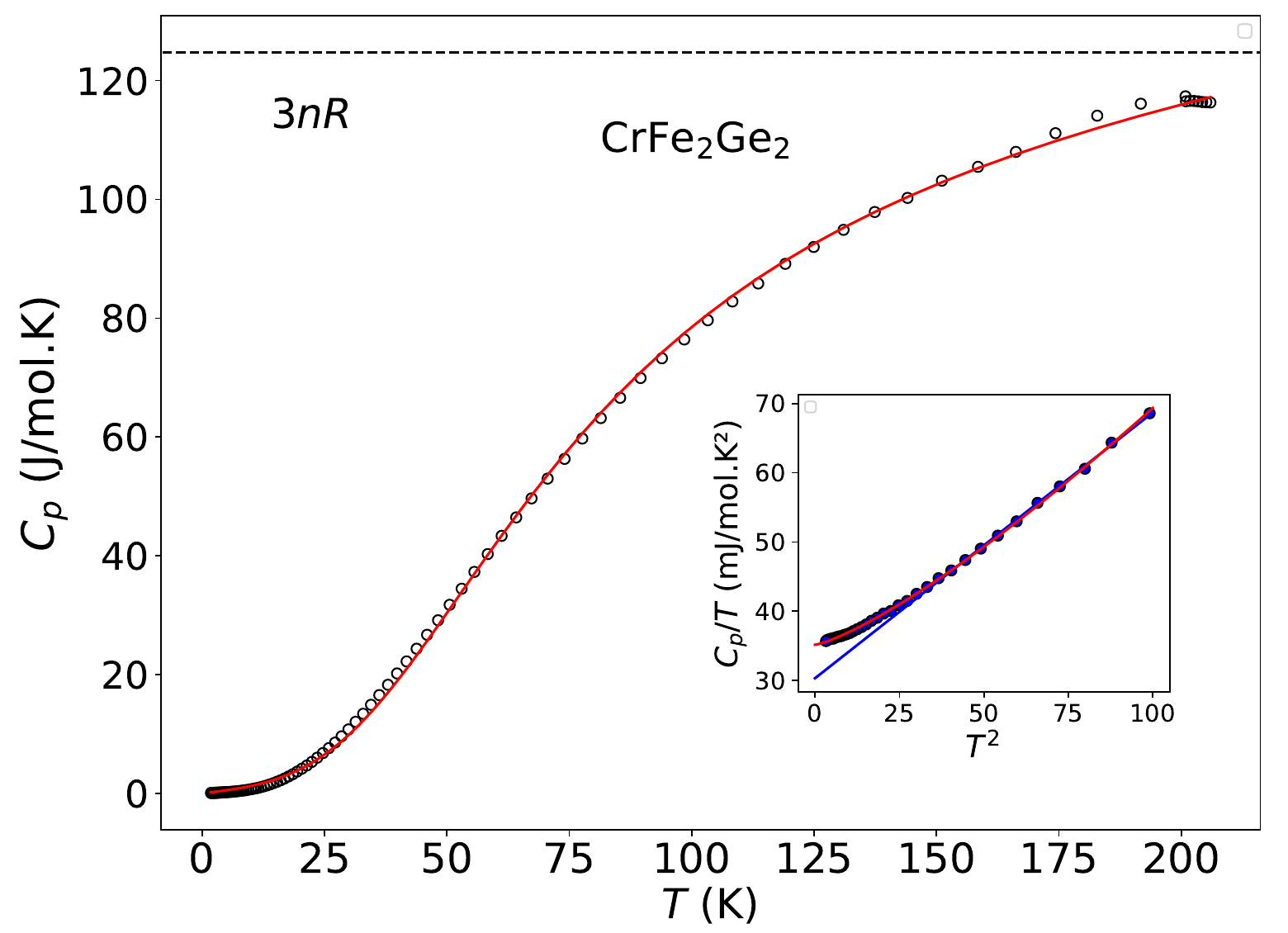}
		\caption{}
	\end{subfigure}%
	
	\caption{(a) Temperature dependence of electrical resistivity (black) and Bloch–Grüneisen fit (red) of $\rm CrFe_2Ge_2$. Inset shows the $\rm < 15~K$ data fitted to $a + bT^n$, $n = 3.06$. (b) Specific heat capacity of $\rm CrFe_2Ge_2$ as a function of
		temperature (black) and the fit to the Debye model (red). The inset shows the measured $T \rm \leq 10~K$, $C_{P}/T$ vs $T^2$ dependence (black) and the fitting to 
		$\gamma + \beta T^{n}$, where $n= 2$ (blue line) and $n=2.273$ (red line). The dashed line represents the high-temperature Dulong-Petit limit  $C_{P} =3nR=\rm 124.72~\, J/mol~K$.}
	\label{fig:transport}
\end{figure*}
The resistivity measurements were performed using a 4-point method in the range $2 - 300~\rm K$ using thin-gauge gold wires and silver paint in the electrical contacts. The resistivity, depicted in Fig. \ref{fig:transport} (a), shows a metallic character throughout the temperature range with a residual resistivity ratio RRR = $\rho_{\rm 300~K}/\rho_{\rm 2~K}=1.6$. No change in slope is observed at $T_{\rm C}$ within the resolution of our data. The thermal dependence of the resistivity follows closely the Bloch-Gr\"{u}neisen law with power 5, typical of metallic systems dominated by scattering of electrons by acoustic phonons,

\begin{equation}
 \rho (T) = \rho_{0} + A \left( \frac{T}{\Theta_{R}
 } \right)^{5} \int_{0}^{\frac{\Theta_{R}}{T}} \frac{x^{5}}{\left( e^x - 1 \right) \left( 1  - e^{-x} \right)}  dx, 
 \label{eq:rho_bg}
\end{equation}
where $\rho_{0}$ is the residual resistivity at $T \rm = 0$, $\Theta_{R}$ corresponds to a characteristic cutoff temperature in the phonon spectrum, usually close to the Debye temperature, and $A$ is a parameter proportional to the electron-phonon coupling. Due to complexity to treat resistivity data from irregular-shaped samples, we limited the use of the relation equation \ref{eq:rho_bg} for the estimation of $\Theta_{R}$. The estimated value for $\Theta_{R}$ is  $\sim \rm 160~K$. At low-temperatures ($<$ 15 K),  a fit to a single power-law $\rho =\rho_0 + r T^n$, yields $n=3.06$, showing deviation from pure electron-electron interactions ($n = 2$) and considerable contribution from electron-magnon ($n=3$) interactions.

The temperature dependence of the specific heat $C_{P}(T)$ of $\rm CrFe_2Ge_2$ is depicted in Fig. \ref{fig:transport} (b).  The specific heat shows a very small hump around $\rm 201~K$. The specific heat was also analyzed in the full measured temperature range using the modified Debye equation: 
\begin{equation}
    C_{D} = \Gamma \cdot T + \alpha 9nR \left( \frac{T}{\Theta_{D}
 } \right)^{3} \int_{0}^{\frac{\Theta_{D}}{T}} \frac{x^{4}e^x}{\left( e^x - 1 \right)^{2}}  dx
 \label{eq:bg_heat}
\end{equation}
where $\Gamma$ accounts for the linear contribution to the specific heat data,  $n$ is the number of atoms per f.u., $R$ is the universal gas constant, $\Theta_{D}$ is the Debye temperature and $\alpha$ was introduced to acquire better fitting at mid-temperature range. The resulting $\alpha $ and $\Theta_{D}$ are $\rm 0.87$ and $\rm 318 ~K$, respectively. The introduction of parameters $\Gamma$ and $\alpha$ improve the fitting while the main parameter, $\theta_D$, varies by less than $\rm 3~\%$.

The low-temperature $\rm < 10~K$ specific heat data are depicted as $C_{P}/T$ versus $T^{2}$ in the inset of Fig. \ref{fig:transport} (b). The data can be fitted by the expression

\begin{equation}
\frac{C_p}{T} = \gamma  + \beta T^{2} 
\label{eq:sommerfeld}\end{equation}
where $\gamma$ is the Sommerfeld electronic specific heat coefficient due to conduction electrons and the second term, $\beta$, is the low-temperature limit of the lattice heat capacity. The coefficient $\beta$ is used for determination  of the Debye temperature relevant for the low-temperature regime.
\begin{equation}
\Theta_D^{*} = \sqrt[3]{\frac{12\pi^{4}nR}{5\beta}}   . 
\end{equation}
 The fitting results are $\gamma = \rm 30~\mathrm{mJ/mol K^{2}}$ and $\beta \rm = 0.38~mJ\,mol^{-1} K^{-4}$ ($\Theta_D^{*} \rm \sim 295~K$). The initial fitting, with equation \ref{eq:sommerfeld} fails to capture the low temperature  interactions apart from electron-phonon interactions. Below $\rm 5 ~K$, $C_{P}/T$ versus $T^{2}$, deviates from linearity. A close to perfect match to experimental data was achieved by fitting without exponent restriction the equation \ref{eq:sommerfeld}, resulting in a modified equation,  $C_P/T=\gamma  + \beta T^{1.27}$ (red line in the Fig. \ref{fig:transport} (b) inset). Resulting in enhancement of  $\gamma = \rm 35.4 ~\mathrm{mJ/mol K^{2}}$. As observed in case of low temperature resistivity behaviour, this enhancement of  $\gamma$ is attributed to electron-magnon interactions. The contribution of magnons to the specific heat would add a $\delta T^{3/2}$ \cite{gopal2012specific} term to the initial equation \ref{eq:sommerfeld}. Accordingly, we have attempted to fit the data to

\begin{equation}
\frac{C_p}{T} = \gamma  + \beta T^{2} + \delta T^{1/2}~.
\label{eq:sommer_modified}\end{equation}

However, the fit of equation (\ref{eq:sommer_modified}) including the magnetic term to $C_P/T$ versus $T^2$ was not successful, yielding either a negative $\delta$ value or a unreasonably small value of $\beta$.   The observed up-turn of the low -$T$ $C_p/T$ versus $T^2$ data could have other origins, including spin wave-stiffness, a non-zero gap \cite{Lees1999} in the magnon spectrum or other undisclosed contributions to low-temperature $C_P$. An inelastic neutron-scattering study would be valuable to clarify this issue.    

\section{Discussion}
\begin{figure*}[htpb]
	\captionsetup{font=small}
	\centering
	\begin{subfigure}{.45\linewidth}
		\centering
		\includegraphics[width=0.95\textwidth]{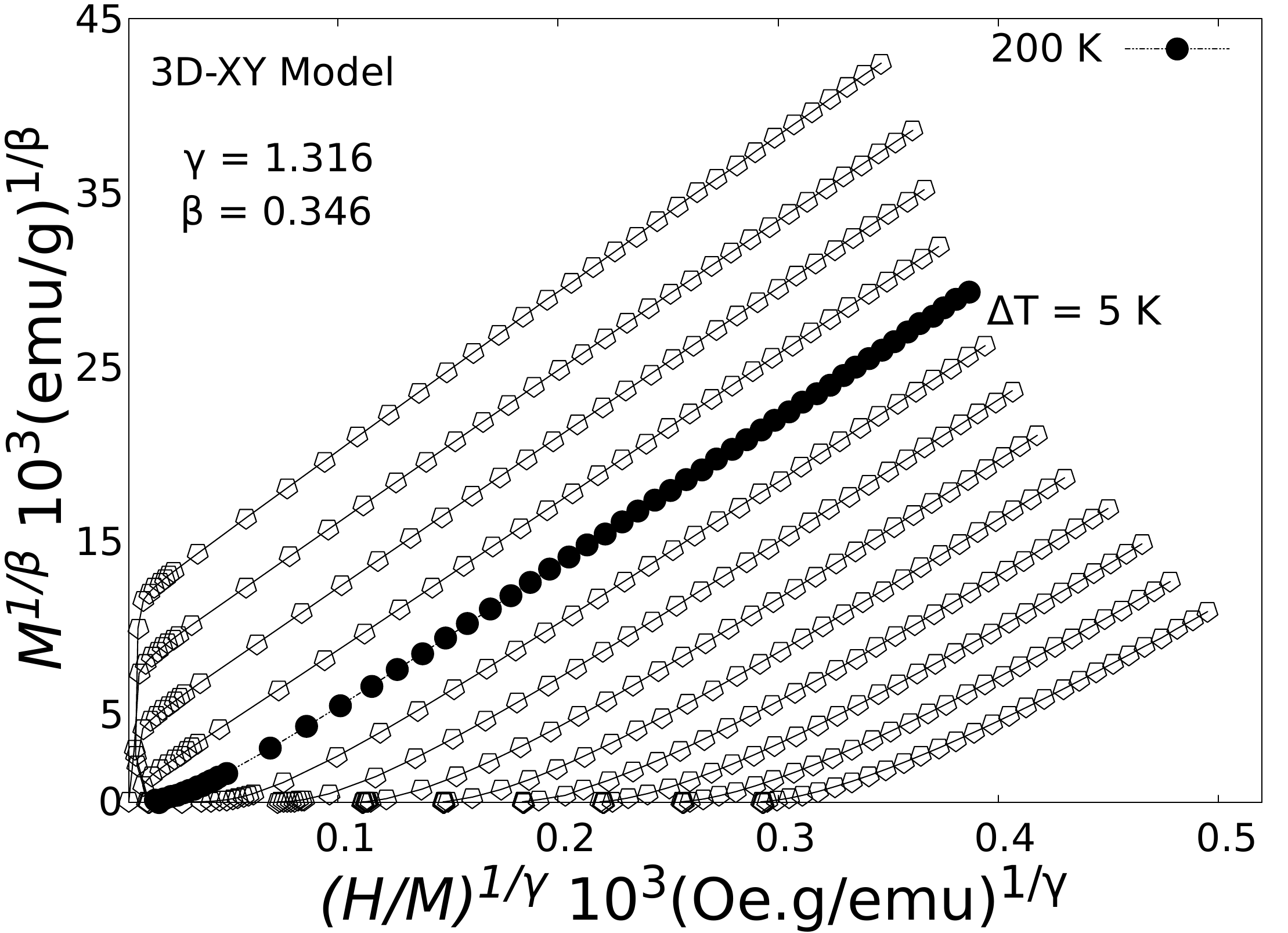}
		\caption{}
		\label{fig:arrot_n2}
	\end{subfigure}%
	\begin{subfigure}{.45\linewidth}
		\centering
		\includegraphics[width = 0.95\linewidth]{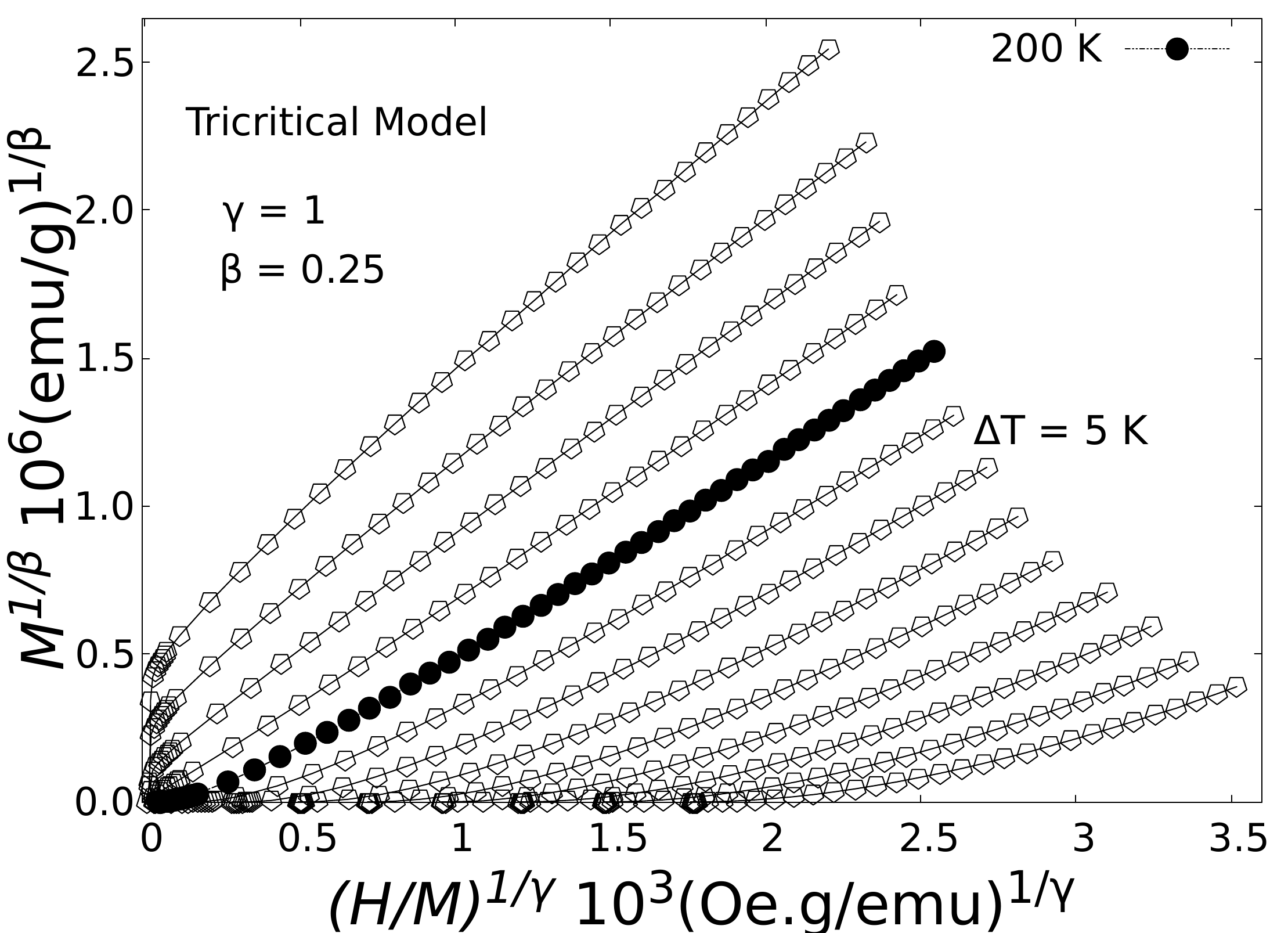}
		\caption{}
		\label{fig:arrot_3crit}
	\end{subfigure}%
	\hspace{0.51em}
	
	\begin{subfigure}{.45\linewidth}
		\centering
		\includegraphics[width = 0.95 \linewidth]{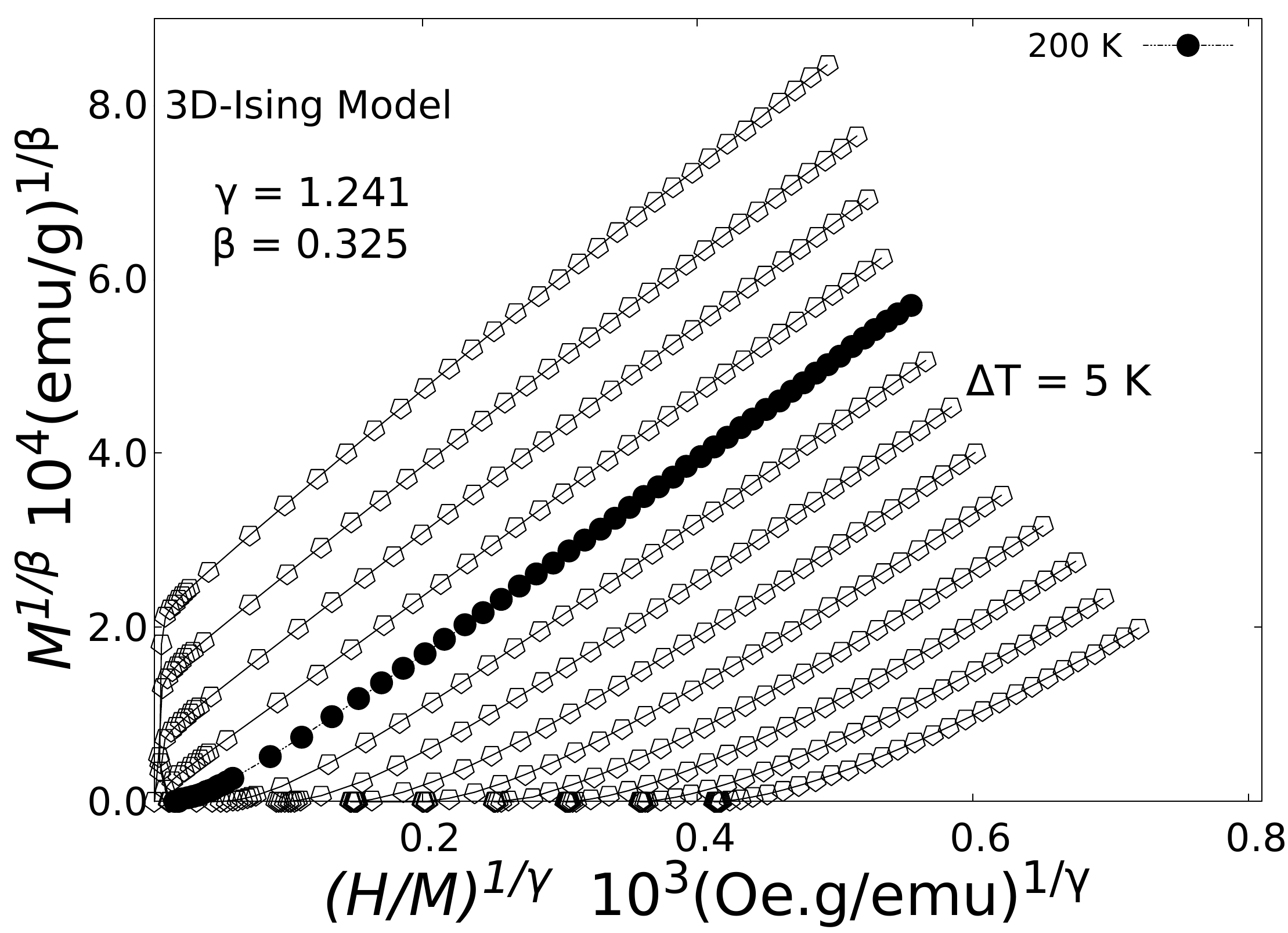}
		\caption{}
		\label{fig:3d_ising}
	\end{subfigure}%
	\hspace{1em}
	\begin{subfigure}{.45\linewidth}
		\centering
		\includegraphics[width =0.95 \linewidth]{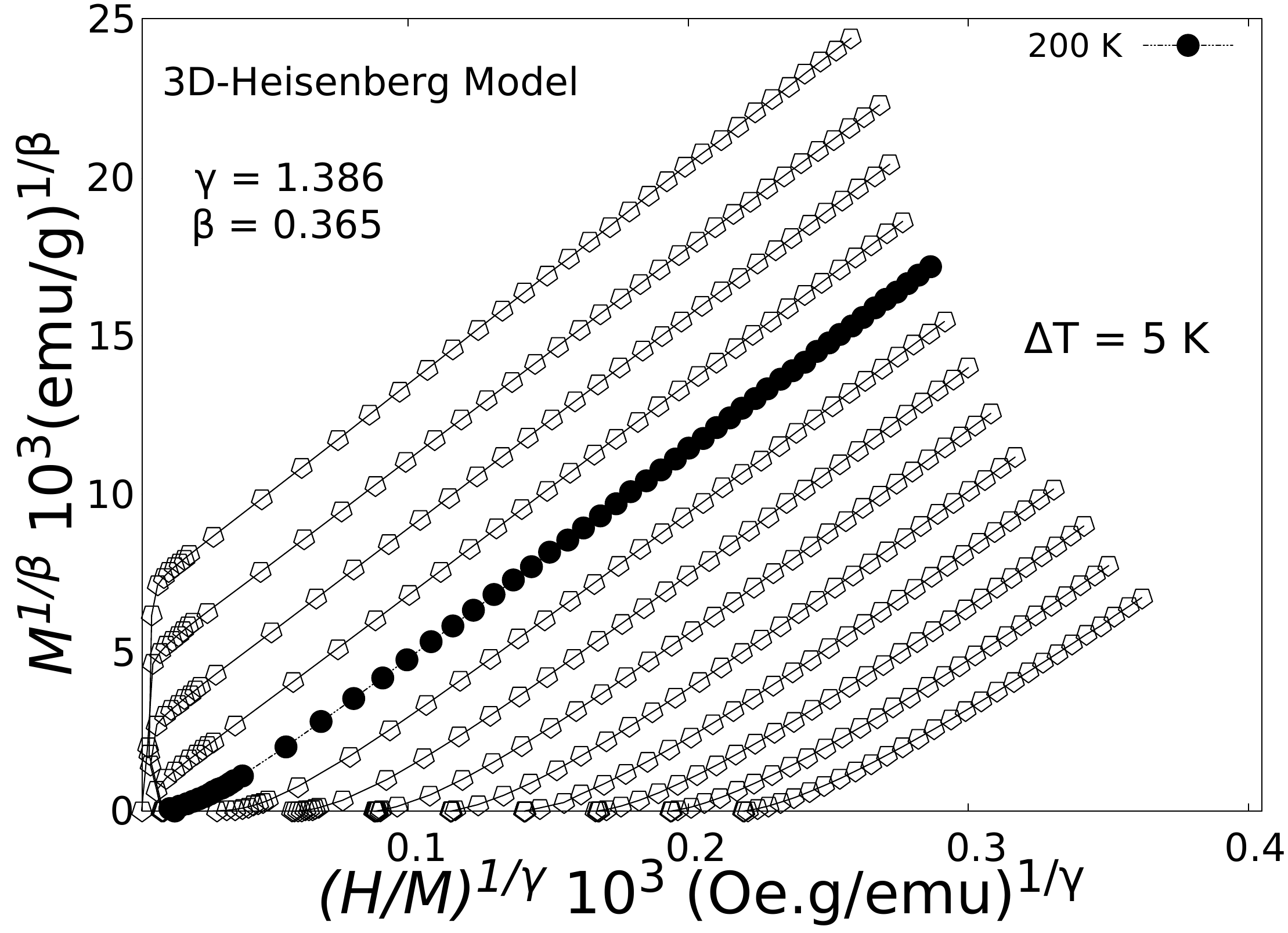}
		\caption{}
		\label{fig:arrot_3d_heisenb}
	\end{subfigure}%
	\hspace{0.51em}
	\begin{subfigure}{.45\linewidth}
		\includegraphics[width = 0.95\linewidth]{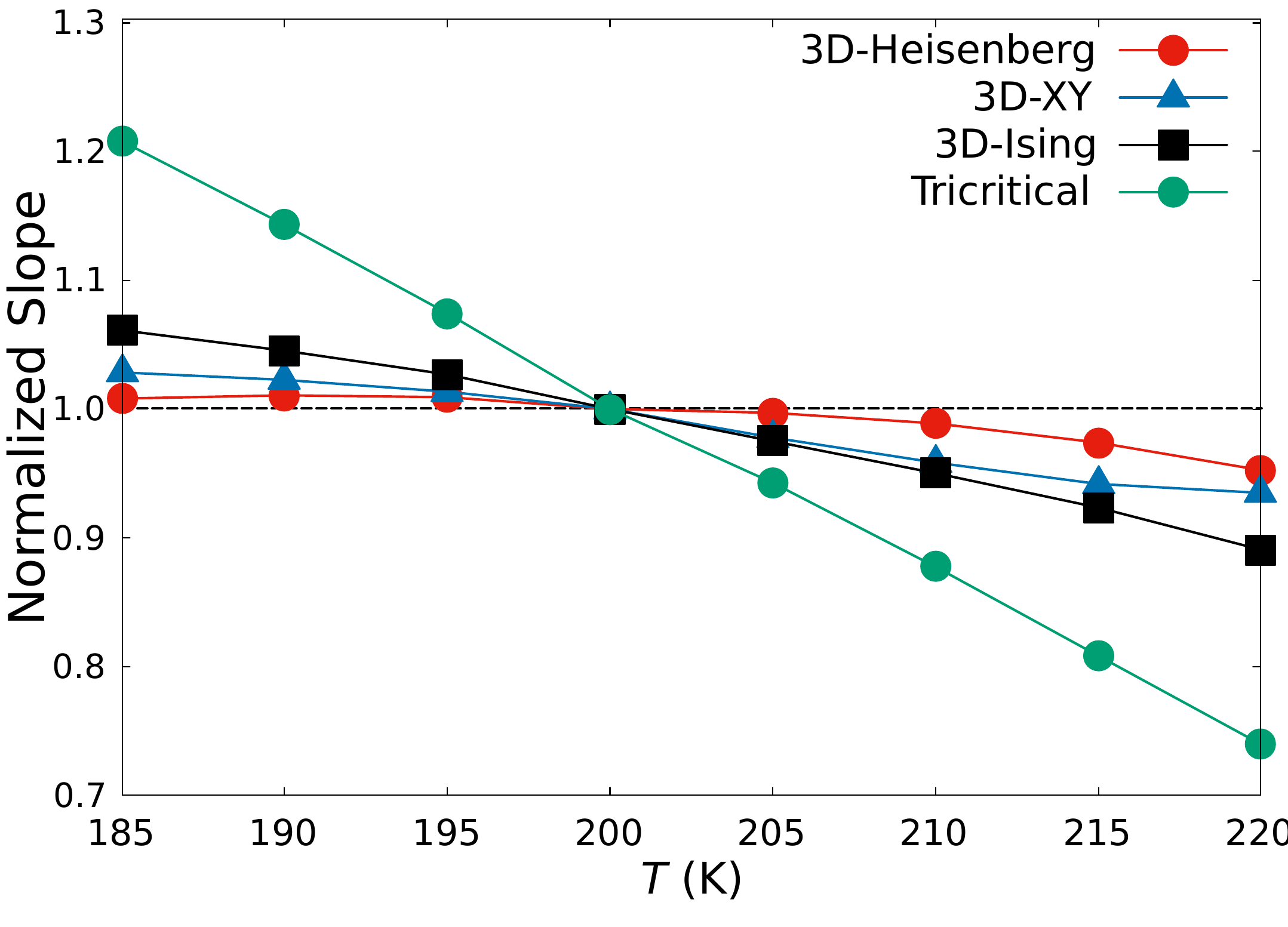}
		\caption{}
		\label{mag:rs2}
	\end{subfigure}%
	\hspace{1em}
	\centering
	\centering
	\begin{subfigure}{.45\linewidth}
		\centering
		\includegraphics[width = 0.95\linewidth]{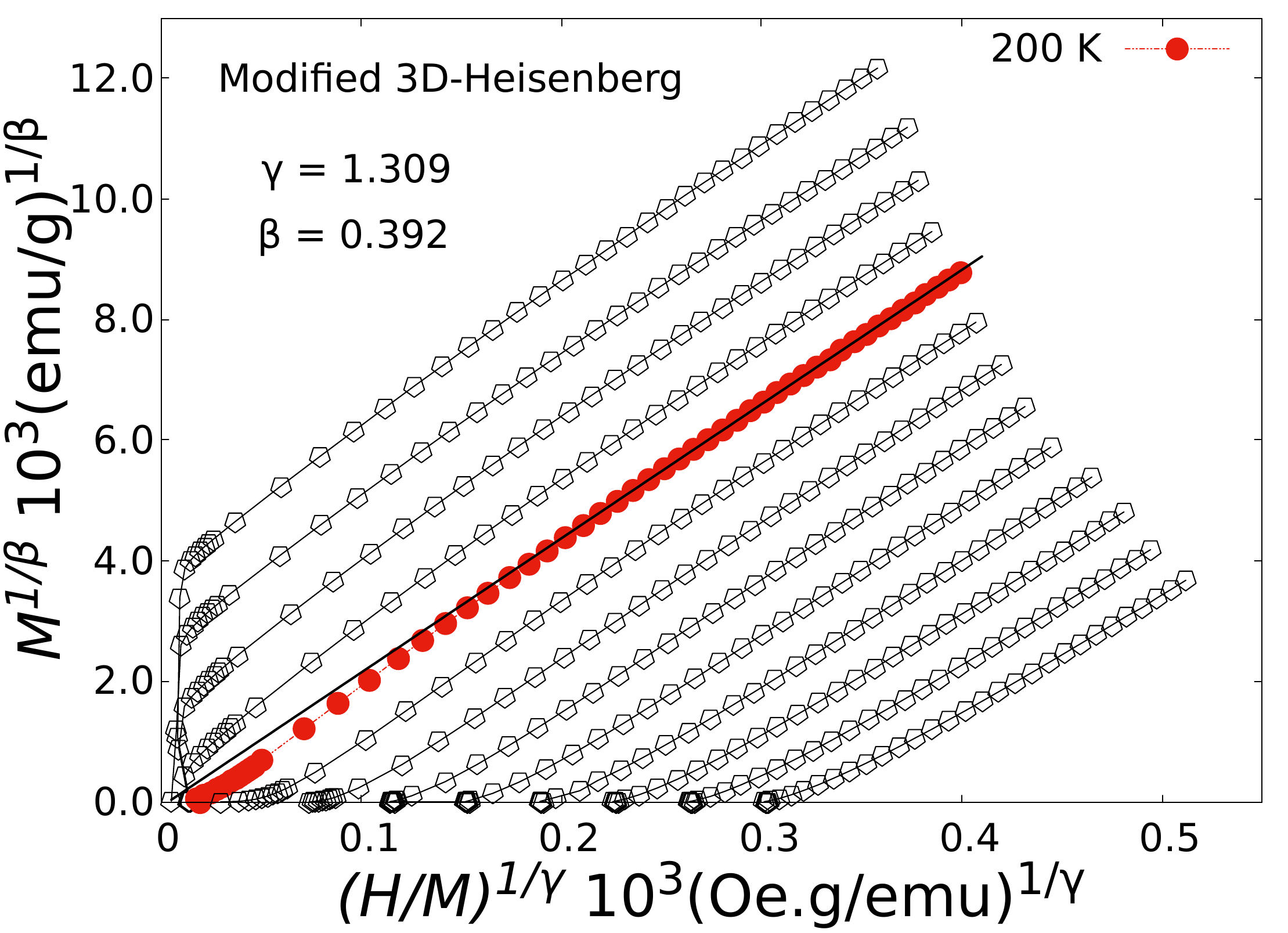}
		\caption{}
		\label{mag:moooo}
	\end{subfigure}%
	
	\caption{(a-d) Isotherms plotted as $M^{1/\beta}$ vs $(H/M)^{1/\gamma}$ with (a) 3D-XY model, (b) Tricritical, (c) 3D-Ising model, and (d) 3D-Heisenberg model. (e) Temperature dependence of normalized slope;   (f) Modified Arrot plot with coefficients $\beta = 0.392$ and $\gamma =$ 1.309 for CrFe$_2$Ge$_2$. The straight line is the linear fit of isotherm at 200\,K
			which passes close to the origin.
	}
	\label{fig:arrots}
\end{figure*}
From the preliminary analyses of bulk magnetization, $\rm CrFe_2Ge_2$ undergoes a second-order thermodynamic phase transition from a ferromagnetic state to a paramagnetic state at a well defined transition temperature $T_{\rm C}$. However, the Landau mean-field theory was not valid, as the isotherm lines of the Arrott plots were not straight and parallel, as demanded from such theory and only a modified Arrott could fit the data. 
In the vicinity of a second-order phase transition, a set of interrelated static critical exponents ($\beta$, $\gamma$, and $\delta$)  characterizes the critical behavior \cite{stanley1971phase}. These exponents are useful to classify the nature of ferromagnetic interactions, the spin dimensionality and their correlation lengths. The mean-field Arrott plot corresponds to the critical exponents $\beta =$ 0.5, $\gamma =$ 1.0 and $\delta = 3$ associated, respectively, with the spontaneous magnetization $M_{\rm s}$ below $T_{\rm C}$, inverse initial magnetic susceptibility $\chi^{-1}$ above $T_{\rm C}$ and critical isotherm magnetization at $T_{\rm C}$, as given by the following expressions \cite{Fisher_1967}:

\begin{equation}
	M_S (T) = M_0 (-\epsilon)^{\beta}, ~\epsilon < 0, ~T < T_C
	\label{eq:beta_ms}\end{equation}

\begin{equation}
		\chi_0^{-1} = (h_0/m_0)\epsilon^{\gamma},~ \epsilon >0,~ T > T_C
	\label{eq:gamma}\end{equation}

\begin{equation}
	M = DH^{1/\delta}, ~\epsilon = 0,~ T = T_C
	\label{eq:isotherm}\end{equation}
where $\epsilon = (T-T_C)/T_C$ is the reduced temperature, $M_0$, $h_0/m_0$ and \textit{D} are the critical amplitudes. These three critical exponents are connected by the Widom relation \cite{Widom1964}

\begin{equation}
	\delta = 1 + \gamma/\beta
	\label{eq:widom}\end{equation}

In case of  $\rm CrFe_2Ge_2$, the modified Arrott plot (MAP) can thus be used to figure out the proper values of critical exponents and find the model that best suits the material. To obtain an appropriate starting point, we first tested the  four three-dimensional models, 3D-Heisenberg ($\beta =$ 0.365, $\gamma = $ 1.386), 3D-XY ($\beta =$ 0.345, $\gamma = $1.316), 3D-Ising model ($\beta =$ 0.325, $\gamma = $ 1.24)\cite{Fisher1974,KAUL19855}, and the tricritical mean-field model ($\beta =$ 0.25, $\gamma = $ 1.0) \cite{KAUL19855}. As shown in Fig. \ref{fig:arrots}, all of them exhibit a set of quasi-straight lines in the high field region, although  the tricritical (3-MF) mode fails to produce paralell lines.

To identify the best model, we compared the relative normalized slope (NS), $\mathrm{NS} = S(T)/S(T_C)$, obtained from the linear fit in the high-field region of the isotherms from each MAP. The best model should give straight and nearly parallel lines in the region of high field; i.e, the MAP normalized slope closer to NS = 1. As shown in Fig.\,\ref{fig:arrots} (e), the 3D Heisenberg model best suits CrFe$_2$Ge$_2$. In an effort to arrive at the most accurate values of $\beta $, $\gamma $ and $T_{\rm C}$, a iteration method was applied, starting from the 3D-Heisenberg model using the following equation of state:

\begin{equation}
	(H/M)^{1/\gamma} = a \left(\frac{T-T_{\rm C}}{T_{\rm C}}\right) + bM^{1/\beta}    
	\label{eq:map_beta}\end{equation}
where \textit{a} and \textit{b} are constants. The outcome of this iterative derivation of exponents gave the following values: $\beta =$ 0.392, $\gamma = $ 1.309, and $\delta = $ 4.34 estimated from the Widom relation. 	Both $\beta$ and $\gamma$ display a shift toward the mean-field values. This shift may result from both electron itinerancy and the influence of magnetic anisotropy in the material. 
\begin{figure*}[h]
	\captionsetup{font=small}
	\centering
	\begin{subfigure}{.45\linewidth}
		\centering
		\includegraphics[width=0.95\linewidth]{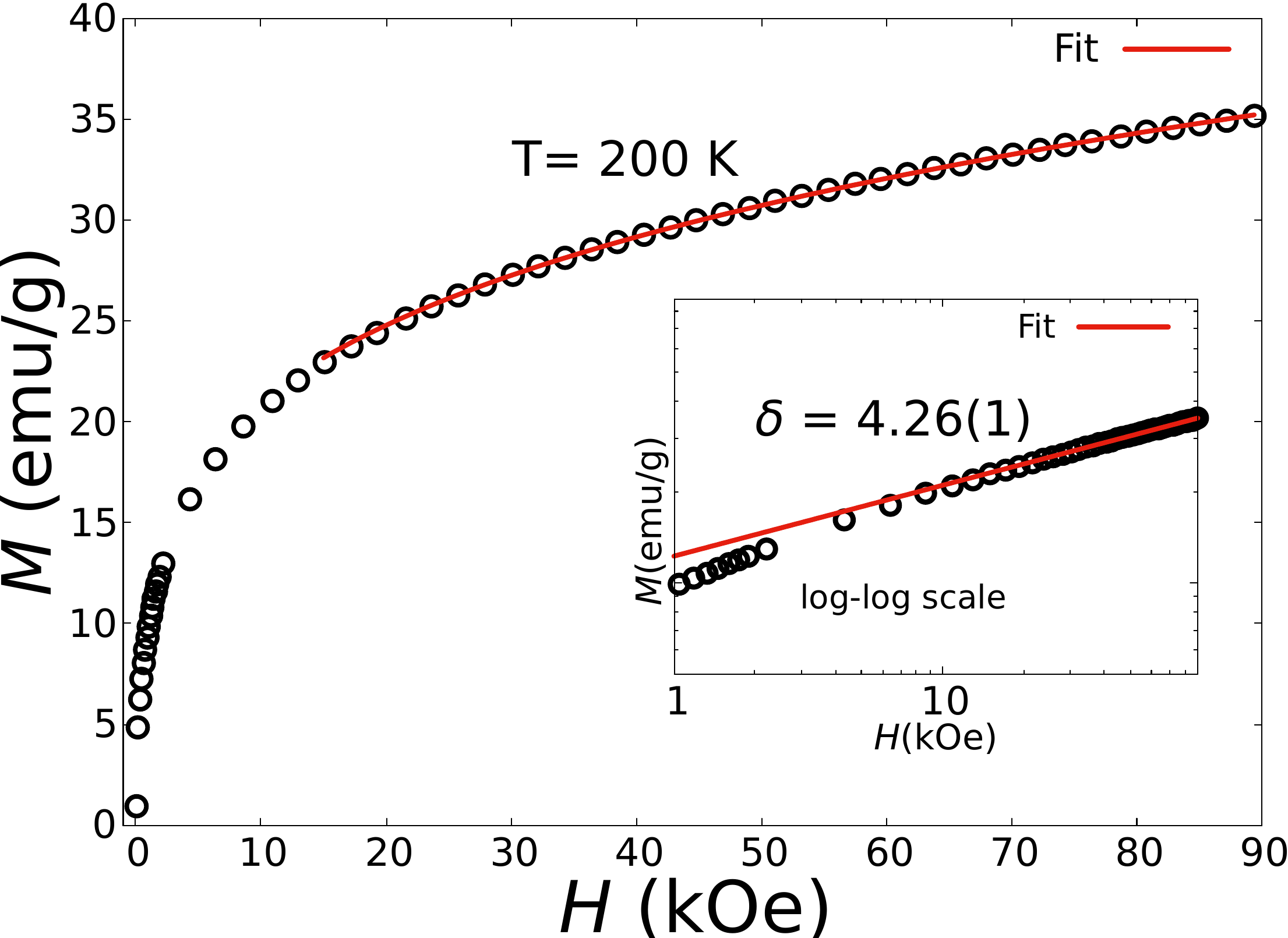}
		\caption{}
		\label{fig:critical_i}
	\end{subfigure}%
	\begin{subfigure}{.45\linewidth}
		\centering
		\includegraphics[width=0.97\linewidth]{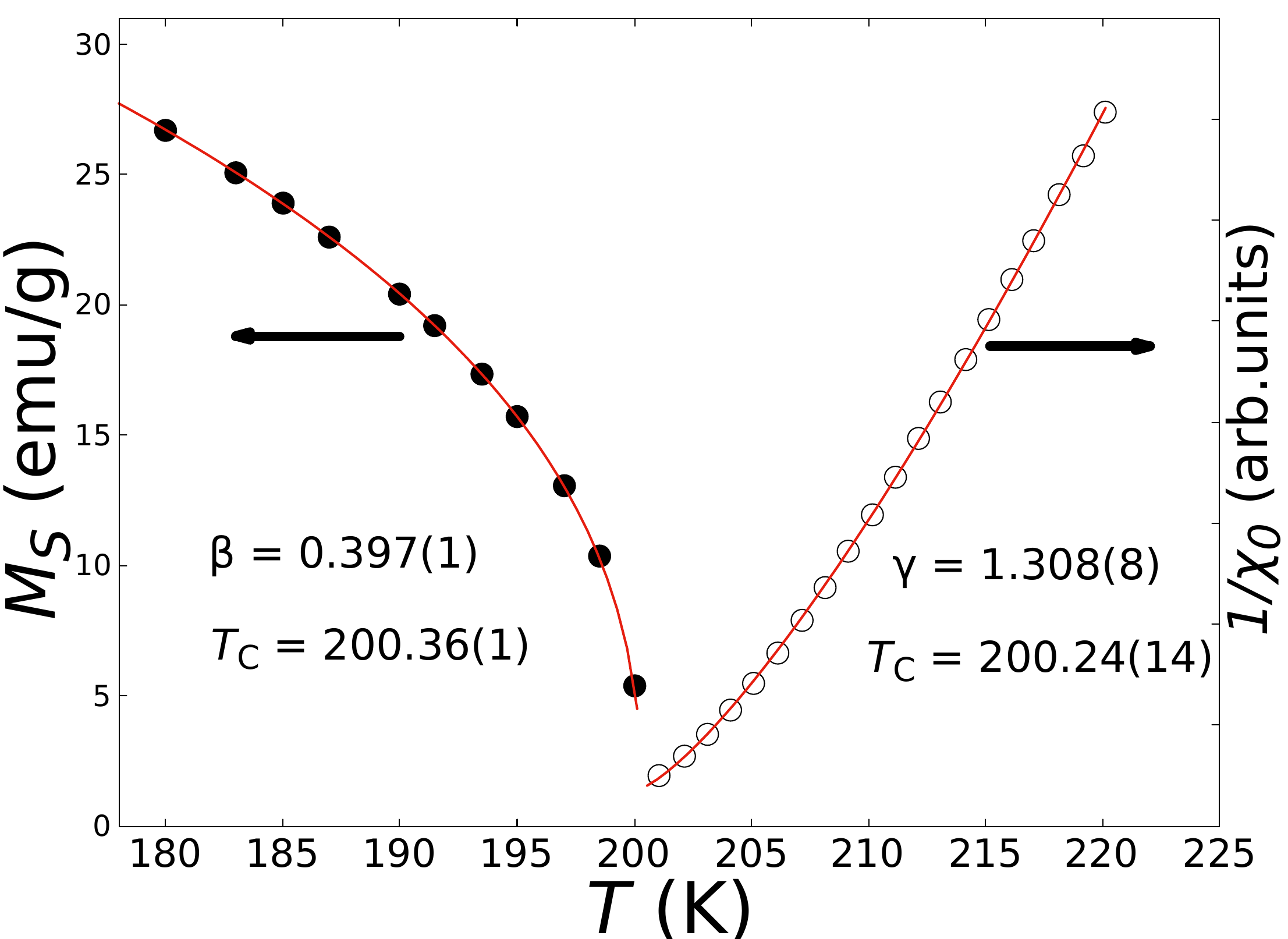}
		\caption{}
		\label{fig:ms_beta}
	\end{subfigure}%
	
	
	\begin{subfigure}{.45\linewidth}
		\centering
		\includegraphics[width =0.95 \linewidth]{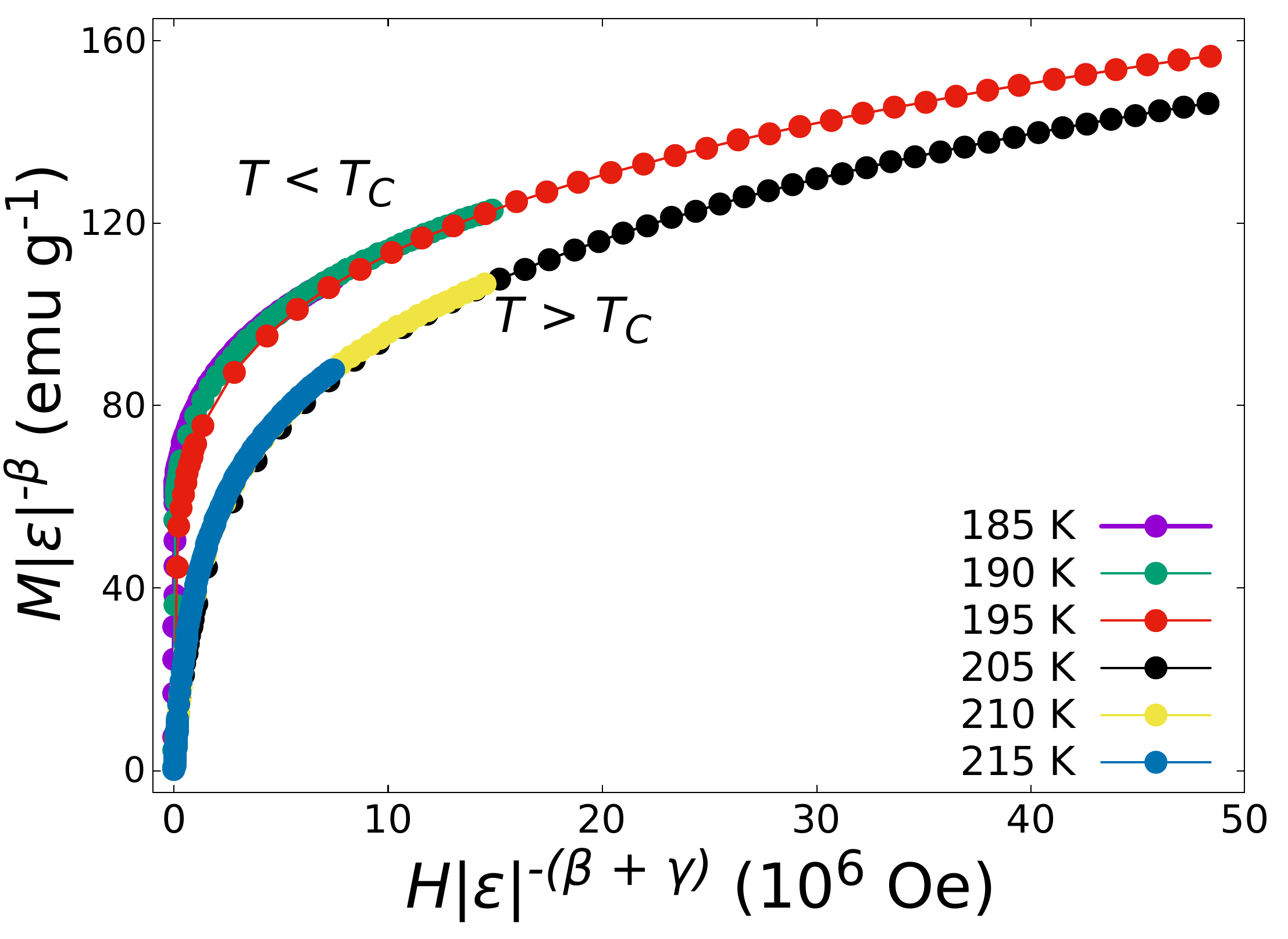}
		\caption{}
		\label{fig:conformit}
	\end{subfigure}%
	
	\caption{ (a) Isotherm M(H) collected at $T_C$ 200 K for  CrFe$_{2}$Ge$_{2}$. Inset shows the same plot in log-log scale and the straight line is the linear fit following Eq.\,(\ref{eq:isotherm}); (b) Temperature dependence of the spontaneous magnetization $M_S$ and $\chi_o^{-1}$ fitted to solid lines from Eq. \ref{eq:beta_ms} and \ref{eq:gamma}, respectively; (c) Scaling plots of renormalized $m$ vs $h$.}
	\label{fig:renormalization}
	
\end{figure*}
Figure \ref{fig:renormalization}\,(a) shows the critical isotherm $M(H)$ at $T= 200\rm ~K$ and in the inset the data is presented in log-log scale, which gives  a straight line of slope  $\delta$. In this particular case, the fitting gives $\delta = 4.26(1)$. The temperature dependence of the obtained $M_S(T)$ and $\chi_{0}^{-1}$ are depicted in Fig. \ref{fig:renormalization}\,(b). By fitting the data of $M_S(T)$ to Eqs. (\ref{eq:beta_ms}) and non-analytical modified (\ref{eq:gamma}) [see Supplementary Information], one obtains two new values  of $\beta = 0.397(1)$ with $T_{\rm C} = 200.36(1)$ and $\gamma =1.308(8)$ with $T_{\rm} = 200.24(14)$, see Fig. \ref{fig:renormalization}\,(b). The calculated $\delta$ from these fitting process is equal 4.29. These results are very close to the value obtained from the modified 3D-Heisenberg Arrott plot of $\gamma = 1.309$, $\beta = 0.392$ and $\delta = 4.34$. Hence, estimated exponents $(\beta ~\mathrm{and} ~\gamma)$ in the present study are self-consistent and reliable.

Finally, these critical exponents should obey the scaling equations. According to the scaling hypothesis, in the asymptotic critical region, the magnetic equation is written as \cite{stanley1971phase}:

\begin{equation}
	M(H,\epsilon) = \epsilon ^{\beta} f_{\pm} (H/ \epsilon ^{\beta+\gamma}) 
	\label{eq:scaling_analysis}
\end{equation}
where $f_{\pm}$ are regular functions denoted as $f_{+}$   for $T > T_C$ and $f_{-}$ for $T < T_C$. The equation (\ref{eq:scaling_analysis}) can be further rewritten in terms of normalized magnetization $m = H |\epsilon|^{-\beta}$ and normalized magnetic field $h = H |\epsilon|^{-(\beta+\gamma)}$ satisfy $m = f _{\pm}(h)$. This demonstrates that scaled $m$ and $h$ will collapse onto two universal curves below $T_C$ and above $T_C$ with proper choice of critical exponents. The plots of scaled  $m$ versus $h$ are obviously separated into two branches below and above $T_C$.  The well-renormalized curves further confirm the reliability of the obtained critical exponents, see Fig. \ref{fig:renormalization}\,(c).

\begin{table*}
	\centering    
		\captionsetup{font=small}
	\caption{Experimental values of the critical exponents for CrFe$_{2}$Ge$_{2}$ and theoretical values of the critical exponents for various models }
	\begin{tabular}[t]{cccccc} \hline
		Materials/Models      &$\{d:n\}$ & Technique  &  $\gamma$ & $\beta$ & $\delta $\\
		\hline
		CrFe$_{2}$Ge$_{2}$ [This work] &   & MAP       &    1.309   & 0.392   &  4.34  \\
$ -$	&	&   C.I.      &    $ -$      &   $ -$     &  4.26 \\
		Mean Field \cite{KAUL19855}    &    $ -$    & Theory    &   1.0      &    0.5  &  3.0   \\ 
		Tricritical \cite{BANERJEE196416} &   $ -$      & Theory    &   1.0      &   0.25  &  5.0   \\
		3D Heisenberg \cite{KAUL19855}   &  $\{3:3\}$      & Theory    &   1.386    & 0.365   &  4.8   \\
		3D Ising \cite{KAUL19855}    &    $\{3:1\}$       & Theory    &   1.241    &  0.325  &  4.82  \\
		3D-XY \cite{KAUL19855}       &$\{3:2\}$          &Theory      & 1.316      & 0.346   & 4.81  \\
		\hline
		\label{tab:crit_exp_tab}
	\end{tabular}
\end{table*}
For a homogeneous magnet, the determination of magnetic exchange interaction \textit{J(r)}, where \textit{r} is the distance of interaction, can help us to understand the spin interactions. Within the framework of the renormalization group theory\,\cite{Fisher_Nickel_1972}, the magnetic exchange decays with the distance \textit{r} in a form 

\begin{equation}
	J(r) \approx r^{-(d+\sigma)}
\end{equation}
where \textit{d} and $\sigma $ are the spatial  dimensionality and a positive constant, respectively. In this work, the spatial dimensionality \textit{d} is considered equal to 3. To further determine the universality class of the magnetic phase transition, we employed the expression\,\cite{Fisher_Nickel_1972}

\begin{eqnarray}
	\gamma = 1 + \frac{4}{d}\frac{n + 2}{n + 8}\Delta\sigma + \frac{8(n + 2)(n - 4)}{d^{2}(n + 8)^{2}} \\ \times \left[1 + \frac{2G(\frac{d}{2})(7n + 20)}{(n - 4)(n + 8)} \right]\Delta\sigma^{2}
	\nonumber \label{eq:rg_frame}
\end{eqnarray}
	where $\Delta\sigma = (\sigma - \frac{d}{2})$, $G(\frac{d}{2}) =3 - \frac{1}{4}(\frac{d}{2})^{2}$, and \textit{n} is the spin dimensionality. The parameter $\sigma$ can be varied, assuming positive values, for a set of values of $\{d:n\}$ to obtain $\gamma = $ 1.309(1), from the tined MAP. According to this model, the spin interaction range is long for $\sigma < $ 2 and short for $\sigma > \mathrm{2}$ \cite{Fisher_Nickel_1972, Fischer_kaul2002}. The remaining exponents can be further extracted via the following scaling relations: $\nu =\gamma/\sigma$, $\alpha = 2 - \nu d$, $\beta  = (2 - \alpha - \gamma)/2$, and $\delta = 1 + \gamma/\beta$. This exercise was performed for different sets of $\{d:n\}$ and the results are resumed in Table \ref{tab:rg_tab}.
	
	\begin{table}    
			\captionsetup{font=small}
		\caption{Critical exponents calculated by the renormalization group theory. }
		\begin{tabular}[t]{cccccc} \hline
			\textit{d}      & \textit{n}     & $\sigma $ & $\gamma$ & $\beta$ & $\delta$ \\
			\hline
			3    & 3      &    1.8684      &   1.309    & 0.396  & 4.30 \\
			3    & 2      &   1.9132      &   1.309    & 0.372  & 4.52  \\
			3    & 1      &    1.9897       &   1.309    & 0.332  & 4.94   \\
			2    & 1      &    1.315       &   1.309    & 0.341  & 4.84   \\
			\hline
			\label{tab:rg_tab}
		\end{tabular}
	\end{table}
	
Using the relations mentioned above, we found that $(\{d:n\}= \{\mathrm{3}:\mathrm{3}\})$ and $\sigma = $ 1.8684 give a set of critical exponents $\beta =$ 0.396, $\gamma =$ 1.309 , and $\delta = $ 4.30, which are close to the critical exponents obtained for the modified 3D-Heisenberg with long-range spin interaction, the interactions decays with distance as $J(r) \approx r ^{-4.86}$. 
\section{Summary}
A new intermetallic compound $\rm CrFe_2Ge_2$ has been synthesized and its structure, magnetic and transport properties were thoroughly investigated. The crystal structure, determined from single-crystal XRD is hexagonal and was found to be related to the $\rm Fe_{13}Ge_8$-type of structure, the Fe atoms occupying two crystallographic sites. 
 M\"{o}ssbauer spectroscopy confirms the existence of these two iron sublattices and different $\rm Fe$ nearest neighbor environments due to distribution of both vacancies and Cr atoms on \textit{6h} sites, which affects the crystal field and magnetic moment of the Fe atoms. On \textit{6g} sites, with the lowest number of \textit{6h} nearest neighbors, Fe atoms seem more sensitive. Approximately 6\,$\%$ show a very low $B_{hf} \sim \rm 2.6~T$ most likely in domains where two or more Cr atoms are replaced by vacancies. The magnetic susceptibility and heat capacity measurements confirm the ferromagnetic ordering with a transition temperature of $\rm 200~K$. The temperature dependence of magnetic susceptibility in the high temperature range follows the modified Curie-Weiss law. The Rhodes–Wohlfarth ratio reveals an itinerant magnetic character.
Critical behavior of magnetism and magnetic interactions  near the PM-FM phase transition have been comprehensively investigated. Reliable critical exponents $(\gamma = $ 1.309, $\beta = $ 0.396 and $\delta = 4.31)$ were obtained and confirmed through the Widom scaling law and scaling equations. These values are further supported by renormalization group calculations, which suggest that the system is close to the isotropic long-range 3D-Heisenberg ferromagnet. Low-temperature measurements of resistivity and specific heat indicate that the electron-magnon interactions are the dominant scattering mechanisms within this temperature range. However, no clear evidence of a magnon contribution was observed in the low-temperature behavior of the specific heat. This unexpected result calls for further studies to resolve the mismatch and clarify the underlying sublattice interactions near $T_{\rm C}$.

\section*{Acknowledgments}P.L.S.C. is grateful to Jos\'{e} Francisco Rodrigues Malta for valuable suggestions regarding intermetallic compound synthesis techniques, and Joao Horta Belo for insightful ideas on magnetic data analysis.  This work was financed through national funds by FCT - Fundação para a Ciência e Tecnologia, I.P. in the framework of the project UID/04564/2025, with DOI identifier 10.54499/UID/04564/2025.
C$^2$TN/IST authors acknowledge the Portuguese Foundation for Science and Technology (FCT), contract UID/04349/2020 and the National Infrastructure Roadmap, LTHMFL-NECL and
LISBOA-01-0145-FEDER-022096.

\section*{Declaration of Competing Interest}
The authors declare that they have no known competing financial interests or personal
relationships that could have appeared to influence the work reported in this paper.

\section*{CRediT authorship contribution statement}

\textbf{P.L.S.C:}~Conceptualization (lead); Formal analysis (equal); Investigation (equal); Methodology (equal); Writing – original draft (equal); Writing – review $\&$ editing (equal). \textbf{B.J.C.V:}~Resources (equal); Formal analysis (equal); Investigation (equal); Writing – original draft (equal); Writing – review $\&$ editing (equal). \textbf{J.C.W:}~Resources (equal); Formal analysis (equal); Investigation (equal); Writing – original draft (equal); Writing – review $\&$ editing (equal). \textbf{P.S.P.S:} Investigation (equal); Writing – review $\&$ editing (equal). \textbf{J.A.P:}~Resources (equal); Formal analysis (equal); Investigation (equal);  Supervision (lead); Writing – review $\&$ editing (equal).
\bibliographystyle{elsarticle-num} 
\bibliography{cas-refs}

@article{BRAUN2019368,
title = {{$\rm CaFe_2Ge_2$} with square-planar iron layers – Closing a gap in the row of {$\rm CaT2Ge2$} (T = {Mn–Zn})},
journal = {Journal of Solid State Chemistry},
volume = {276},
pages = {368-375},
year = {2019},
issn = {0022-4596},
doi = {https://doi.org/10.1016/j.jssc.2019.05.032},
url = {https://www.sciencedirect.com/science/article/pii/S0022459619302610},
author = {Thomas Braun and Viktor Hlukhyy}
}

@article{WELTER200335,
title = {Magnetic behaviour of the Mn sublattice in {ThCr2Si2}-type {CaMn2−xFexGe2} solid solution investigated by magnetic measurements and neutron diffraction},
journal = {Journal of Alloys and Compounds},
volume = {354},
number = {1},
pages = {35-46},
year = {2003},
issn = {0925-8388},
doi = {https://doi.org/10.1016/S0925-8388(02)01354-3},
url = {https://www.sciencedirect.com/science/article/pii/S0925838802013543},
author = {R. Welter and B. Malaman}
}

@article{EBIHARA1995219,
title = {Magnetic properties of single crystal {CeFe2Ge2}},
journal = {Physica B: Condensed Matter},
volume = {206-207},
pages = {219-221},
year = {1995},
note = {Proceedings of the International Conference on Strongly Correlated Electron Systems},
issn = {0921-4526},
doi = {https://doi.org/10.1016/0921-4526(94)00413-P},
url = {https://www.sciencedirect.com/science/article/pii/092145269400413P},
author = {T. Ebihara and K. Motoki and H. Toshima and M. Takashita and N. Kimura and H. Sugawara and K. Ichihashi and R. Settai and Y. Ōnuki and Y. Aoki and H. Sato}
}

@article{c,
  title = {Composition dependence of bulk superconductivity in ${\mathrm{YFe}}_{2}{\mathrm{Ge}}_{2}$},
  author = {Chen, Jiasheng and Gam\ifmmode \dot{z}\else \.{z}\fi{}a, Monika B. and Semeniuk, Konstantin and Grosche, F. Malte},
  journal = {Phys. Rev. B},
  volume = {99},
  issue = {2},
  pages = {020501},
  numpages = {5},
  year = {2019},
  month = {Jan},
  publisher = {American Physical Society},
  doi = {10.1103/PhysRevB.99.020501},
  url = {https://link.aps.org/doi/10.1103/PhysRevB.99.020501}
}

@article{Jiang2025,
	author = {Jiang, Jian and Zhang, Xiaolin and Wang, Hao and Yin, Lei and Wen, Yao and Cheng, Ruiqing and Zhang, Chendong and He, Jun},
	title = {Epitaxial Strain Engineering for High-Temperature Ferromagnetic Iron Germanide Alloy},
	journal = {Nano Letters},
	volume = {25},
	number = {24},
	pages = {9639-9645},
	year = {2025},
	doi = {10.1021/acs.nanolett.5c01388},
	note ={PMID: 40461408},
	URL = {https://doi.org/10.1021/acs.nanolett.5c01388},
	eprint = {https://doi.org/10.1021/acs.nanolett.5c01388}
}

@article{AVILA200451,
title = {Anisotropic magnetization, specific heat and resistivity of {RFe2Ge2} single crystals},
journal = {Journal of Magnetism and Magnetic Materials},
volume = {270},
number = {1},
pages = {51-76},
year = {2004},
issn = {0304-8853},
doi = {https://doi.org/10.1016/S0304-8853(03)00672-3},
url = {https://www.sciencedirect.com/science/article/pii/S0304885303006723},
author = {M.A. Avila and S.L. Bud'ko and P.C. Canfield}
}

@article{may2016competing,
  title={Competing magnetic ground states and their coupling to the crystal lattice in {CuFe2Ge2}},
  author={May, Andrew F and Calder, Stuart and Parker, David S and Sales, Brian C and McGuire, Michael A},
  journal={Scientific reports},
  volume={6},
  number={1},
  pages={35325},
  doi={https://doi.org/10.1038/srep35325},
  year={2016},
  publisher={Nature Publishing Group UK London}
}

@article{HARRIS1989103,
title = {Structural, magnetic and constitutional studies of a new family of ternary phases based on the compound {Fe3GaAs}},
journal = {Journal of the Less Common Metals},
volume = {146},
pages = {103-119},
year = {1989},
issn = {0022-5088},
doi = {https://doi.org/10.1016/0022-5088(89)90367-6},
url = {https://www.sciencedirect.com/science/article/pii/0022508889903676},
author = {I.R. Harris and N.A. Smith and E. Devlin and B. Cockayne and W.R. Macewan and G. Longworth}
}

@article{MALAMAN1980155,
title = {Étude structurale des germaniures {Fe(Co)2−xGe} de type $\beta$ et $\eta$, et de leurs alliages avec le gallium {Fe(Co)2−xGe1−yGay}},
journal = {Journal of the Less Common Metals},
volume = {75},
number = {2},
pages = {155-176},
year = {1980},
issn = {0022-5088},
doi = {https://doi.org/10.1016/0022-5088(80)90114-9},
url = {https://www.sciencedirect.com/science/article/pii/0022508880901149},
author = {B Malaman and J Steinmetz and B Roques}
}

@article{OMoze_1994,
doi = {10.1088/0953-8984/6/48/005},
url = {https://dx.doi.org/10.1088/0953-8984/6/48/005},
year = {1994},
month = {nov},
publisher = {},
volume = {6},
number = {48},
pages = {10435},
author = {O Moze and  C Greaves and  F Bouree-Vigneron and  B Cockayne and  W R MacEwan and  N A Smith and  I R Harris},
title = {Neutron power diffraction study of {Fe3Ga1.7As0.3} and {Fe3Ga1.15Sb0.85} B82-type compounds},
journal = {Journal of Physics: Condensed Matter}
}

@article{ADELSON19651795,
title = {Magnetic structures of iron germanides},
journal = {Journal of Physics and Chemistry of Solids},
volume = {26},
number = {12},
pages = {1795-1804},
year = {1965},
issn = {0022-3697},
doi = {https://doi.org/10.1016/0022-3697(65)90212-X},
url = {https://www.sciencedirect.com/science/article/pii/002236976590212X},
author = {E. Adelson and A.E. Austin}
}

@article{Bei_Ding2020,
author = {Ding, Bei and Li, Zefang and Xu, Guizhou and Li, Hang and Hou, Zhipeng and Liu, Enke and Xi, Xuekui and Xu, Feng and Yao, Yuan and Wang, Wenhong},
title = {Observation of Magnetic Skyrmion Bubbles in a van der Waals Ferromagnet Fe3GeTe2},
journal = {Nano Letters},
volume = {20},
number = {2},
pages = {868-873},
year = {2020},
doi = {10.1021/acs.nanolett.9b03453},
note ={PMID: 31869236},
URL = {https://doi.org/10.1021/acs.nanolett.9b03453},
eprint = {https://doi.org/10.1021/acs.nanolett.9b03453}
}

@article{Liu2016-pz,
  title    = "Critical behavior of the quasi-two-dimensional semiconducting
              ferromagnet {CrSiTe3}",
  author   = "Liu, Bingjie and Zou, Youming and Zhang, Lei and Zhou, Shiming
              and Wang, Zhe and Wang, Weike and Qu, Zhe and Zhang, Yuheng",
  journal  = "Scientific Reports",
  volume   =  6,
  number   =  1,
  pages    = "33873",
  month    =  sep,
  year     =  2016
}

@article{Albertini1998,
    author = {Albertini, F. and Pareti, L. and Deriu, A. and Negri, D. and Calestani, G. and Moze, O. and Kennedy, S. J. and Sonntag, R.},
    title = {A magnetic and structural study of {Mn}, {Co}, and {Ni} substituted {Fe3Ge2} hexagonal germanides},
    journal = {Journal of Applied Physics},
    volume = {84},
    number = {1},
    pages = {401-410},
    year = {1998},
    month = {07},
    issn = {0021-8979},
    doi = {10.1063/1.368080},
    url = {https://doi.org/10.1063/1.368080}
}

@article{Kitagawa2022,
author = {Kitagawa ,Jiro},
title = {Magnetic Properties and Magnetocaloric Effect of {Fe3Ga0.35Ge1.65}},
journal = {Journal of the Physical Society of Japan},
volume = {91},
number = {6},
pages = {065004},
year = {2022},
doi = {10.7566/JPSJ.91.065004},
URL = {https://doi.org/10.7566/JPSJ.91.065004},
eprint = {https://doi.org/10.7566/JPSJ.91.065004}
}

@article{Doebelin:kc5013,
author = "Doebelin, Nicola and Kleeberg, Reinhard",
title = "{{\it Profex}: a graphical user interface for the Rietveld refinement program {\it BGMN}}",
journal = "Journal of Applied Crystallography",
year = "2015",
volume = "48",
number = "5",
pages = "1573--1580",
month = "Oct",
doi = {10.1107/S1600576715014685},
url = {https://doi.org/10.1107/S1600576715014685}
}

@article{sheldrick2015crystal,
  title={Crystal structure refinement with {SHELXL}},
  author={Sheldrick, George M},
  journal={Acta Crystallographica Section C: Structural Chemistry},
  volume={71},
  number={1},
  pages={3--8},
  year={2015},
  doi={https://journals.iucr.org/c/issues/2015/01/00/fa3356/fa3356.pdf},
  publisher={International Union of Crystallography}
}

@article{zavalij1987structure,
  title={Structure cristalline des compos{\'e}s $\rm CuFe2Ge2$ et $\rm Cu_{1\pm x}Co_{2\pm x}Ge_2$},
  author={ZAVALIJ, I and PECHARSKIJ, VK and BODAK, OI},
  journal={Kristallografi{\^a}},
  volume={32},
  number={1},
  pages={66--69},
  year={1987}
}

@article{BUDKO2018260,
	title = {On magnetic structure of CuFe2Ge2: Constrains from the 57Fe Mössbauer spectroscopy},
	journal = {Journal of Magnetism and Magnetic Materials},
	volume = {446},
	pages = {260-263},
	year = {2018},
	issn = {0304-8853},
	doi = {https://doi.org/10.1016/j.jmmm.2017.09.046},
	url = {https://www.sciencedirect.com/science/article/pii/S0304885317319315},
	author = {Sergey L. Bud’ko and Na Hyun Jo and Savannah S. Downing and Paul C. Canfield}
}

@article{Qin2011,
	title = {Homogeneous amorphous Fe${}_{x}{\mathrm{Ge}}_{1\ensuremath{-}x}$ magnetic semiconductor films with high Curie temperature and high magnetization},
	author = {Qin, Yu-feng and Yan, Shi-shen and Kang, Shi-shou and Xiao, Shu-qin and Zhang, Qiong and Yao, Xin-xin and Xu, Tong-shuai and Tian, Yu-feng and Dai, You-yong and Liu, Guo-lei and Chen, Yan-xue and Mei, Liang-mo and Ji, Gang and Zhang, Ze},
	journal = {Phys. Rev. B},
	volume = {83},
	issue = {23},
	pages = {235214},
	numpages = {7},
	year = {2011},
	month = {Jun},
	publisher = {American Physical Society},
	doi = {10.1103/PhysRevB.83.235214},
	url = {https://link.aps.org/doi/10.1103/PhysRevB.83.235214}
}

@article{Deng2018,
	title    = {Gate-tunable room-temperature ferromagnetism in two-dimensional
	{Fe3GeTe2}},
	author   = {Deng, Yujun and Yu, Yijun and Song, Yichen and Zhang, Jingzhao
	and Wang, Nai Zhou and Sun, Zeyuan and Yi, Yangfan and Wu, Yi
	Zheng and Wu, Shiwei and Zhu, Junyi and Wang, Jing and Chen, Xian
	Hui and Zhang, Yuanbo},
	journal  = {Nature},
	volume   =  {563},
	number   =  {7729},
	pages    = {94--99},
	month    =  {nov},
	doi = {https://doi.org/10.1038/s41586-018-0626-9},
	year     = {2018}
}

@article{Lidin:an0539,
author = {Lidin, S.},
title = {{Superstructure Ordering of Intermetallics: B8 Structures in the Pseudo-Cubic Regime}},
journal = {{Acta Crystallographica Section B}},
year = {1998},
volume = {54},
number = {2},
pages = {97--108},
month = {Apr},
doi = {10.1107/S010876819701879X},
url = {https://doi.org/10.1107/S010876819701879X}
}

@article{Zou2014yfe2ge2,
author = {Zou, Y. and Feng, Z. and Logg, P. W. and Chen, J. and Lampronti, G. and Grosche, F. M.},
title = {Fermi liquid breakdown and evidence for superconductivity in YFe2Ge2},
journal = {physica status solidi (RRL) – Rapid Research Letters},
volume = {8},
number = {11},
pages = {928-930},
doi = {https://doi.org/10.1002/pssr.201409418},
url = {https://onlinelibrary.wiley.com/doi/abs/10.1002/pssr.201409418},
eprint = {https://onlinelibrary.wiley.com/doi/pdf/10.1002/pssr.201409418},
year = {2014}
}

@article{Yu_Liu2020,
  title = {Three-dimensional Ising ferrimagnetism of Cr-Fe-Cr trimers in $\mathrm{Fe}{\mathrm{Cr}}_{2}{\mathrm{Te}}_{4}$},
  author = {Liu, Yu and Koch, R. J. and Hu, Zhixiang and Aryal, Niraj and Stavitski, Eli and Tong, Xiao and Attenkofer, Klaus and Bozin, E. S. and Yin, Weiguo and Petrovic, C.},
  journal = {Phys. Rev. B},
  volume = {102},
  issue = {8},
  pages = {085158},
  numpages = {7},
  year = {2020},
  month = {Aug},
  publisher = {American Physical Society},
  doi = {10.1103/PhysRevB.102.085158},
  url = {https://link.aps.org/doi/10.1103/PhysRevB.102.085158}
}

@article{Dara_2023,
doi = {10.1088/1361-648X/ace0ac},
url = {https://dx.doi.org/10.1088/1361-648X/ace0ac},
year = {2023},
month = {jun},
publisher = {IOP Publishing},
volume = {35},
number = {39},
pages = {395802},
author = {Dara, Hanuma Kumar and Patra, Debashish and Moharana, Gyanti Prakash and Sarangi, S N and Samal, D},
title = {Evidence of weak itinerant ferromagnetism and Griffiths like phase in MnFeGe},
journal = {Journal of Physics: Condensed Matter}
}

@article{V_Carteaux_1995,
doi = {10.1209/0295-5075/29/3/011},
url = {https://dx.doi.org/10.1209/0295-5075/29/3/011},
year = {1995},
month = {jan},
publisher = {},
volume = {29},
number = {3},
pages = {251},
author = {V. Carteaux and F. Moussa and M. Spiesser},
title = {2D Ising-Like Ferromagnetic Behaviour for the Lamellar Cr2Si2Te6 Compound: A Neutron Scattering Investigation},
journal = {Europhysics Letters}
}

@article{G_Lin2017,
  title = {{Tricritical behavior of the two-dimensional intrinsically ferromagnetic semiconductor ${\mathbf{CrGeTe}}_{3}$}},
  author = {Lin, G. T. and Zhuang, H. L. and Luo, X. and Liu, B. J. and Chen, F. C. and Yan, J. and Sun, Y. and Zhou, J. and Lu, W. J. and Tong, P. and Sheng, Z. G. and Qu, Z. and Song, W. H. and Zhu, X. B. and Sun, Y. P.},
  journal = {Phys. Rev. B},
  volume = {95},
  issue = {24},
  pages = {245212},
  numpages = {7},
  year = {2017},
  month = {Jun},
  publisher = {American Physical Society},
  doi = {10.1103/PhysRevB.95.245212},
  url = {https://link.aps.org/doi/10.1103/PhysRevB.95.245212}
}

@article{chen2013magnetic,
  title={Magnetic properties of layered itinerant electron ferromagnet {$\rm Fe_3GeTe_2$}},
  author={Chen, Bin and Yang, JinHu and Wang, HangDong and Imai, Masaki and Ohta, Hiroto and Michioka, Chishiro and Yoshimura, Kazuyoshi and Fang, MingHu},
  journal={Journal of the Physical Society of Japan},
  volume={82},
  number={12},
  pages={124711},
  year={2013},
  doi = {https://journals.jps.jp/doi/10.7566/JPSJ.82.124711},
  publisher={The Physical Society of Japan}
}

@article{Jiawang_Xu2024,
author = {Xu, Jiawang and Xi, Lei and Xing, Shouyuan and Sheng, Junchao and Li, Shihao and Wang, Liming and Kan, Xucai and Ma, Tianping and Zang, Yipeng and Bao, Bin and Zhou, Zhonghao and Yang, Mengmeng and Gao, Yawei and Wang, Dingsong and Wang, Guyue and Zheng, Xinqi and Zhang, Jingyan and Du, Haifeng and Xu, Juping and Yin, Wen and Zhang, Ying and Zhou, Shiming and Shen, Baogen and Wang, Shouguo},
title = {Magnetic Structure-Dependent Spin Texture Lattice in Hexagonal MnFeCoGe Magnets},
journal = {ACS Nano},
volume = {18},
number = {35},
pages = {24515-24522},
year = {2024},
doi = {10.1021/acsnano.4c08703},
note ={PMID: 39165001},
URL = {https://doi.org/10.1021/acsnano.4c08703},
eprint = {https://doi.org/10.1021/acsnano.4c08703}
}

@article{Yu2011-ph,
	title    = "Near room-temperature formation of a skyrmion crystal in
	thin-films of the helimagnet {FeGe}",
	author   = "Yu, X Z and Kanazawa, N and Onose, Y and Kimoto, K and Zhang, W Z
	and Ishiwata, S and Matsui, Y and Tokura, Y",
	journal  = "Nature Materials",
	volume   =  10,
	number   =  2,
	pages    = "106--109",
	month    =  feb,
	year     =  2011
}

@article{Maat2008,
	author = {Maat, S. and Carey, M. J. and Childress, J. R.},
	title = {Current perpendicular to the plane spin-valves with CoFeGe magnetic layers},
	journal = {Applied Physics Letters},
	volume = {93},
	number = {14},
	pages = {143505},
	year = {2008},
	month = {10},
	doi = {10.1063/1.2993213},
	url = {https://doi.org/10.1063/1.2993213},
	eprint = {https://pubs.aip.org/aip/apl/article-pdf/doi/10.1063/1.2993213/13316163/143505_1_online.pdf},
}

@article{long1983ideal,
  title={The ideal M{\"o}ssbauer effect absorber thickness},
  author={Long, Gary J and Cranshaw, TE and Longworth, G},
  journal={M{\"o}ssbauer effect reference and data journal},
  volume={6},
  number={2},
  pages={42--49},
  year={1983}
}

@article{GREAVES1990315,
title = {Structural identification of the new magnetic phases {Fe$_3$Ga$_{2 − x}$As$_x$}},
journal = {Journal of the Less Common Metals},
volume = {157},
number = {2},
pages = {315-325},
year = {1990},
issn = {0022-5088},
doi = {https://doi.org/10.1016/0022-5088(90)90187-O},
url = {https://www.sciencedirect.com/science/article/pii/002250889090187O},
author = {C Greaves and E.J Devlin and N.A Smith and I.R Harris and B Cockayne and W.R MacEwan},
}

@article{rhodes1963effective,
  title={The effective Curie-Weiss constant of ferromagnetic metals and alloys},
  author={Rhodes, P and Wohlfarth, Eo P},
  journal={Proceedings of the Royal Society of London. Series A. Mathematical and Physical Sciences},
  volume={273},
  number={1353},
  pages={247--258},
  year={1963},
  doi={https://doi.org/10.1098/rspa.1963.0086},
  publisher={The Royal Society London}
}

@article{shanavas2015,
    doi = {10.1371/journal.pone.0121186},
    author = {Shanavas, K. V. AND Singh, David J.},
    journal = {PLOS ONE},
    publisher = {Public Library of Science},
    title = {{Itinerant Magnetism in Metallic CuFe2Ge2}},
    year = {2015},
    month = {03},
    volume = {10},
    url = {https://doi.org/10.1371/journal.pone.0121186},
    pages = {1-9},
    number = {3}
}

@article{TENER2021167827,
title = {Magnetization distribution in {$\rm Cu_{0.6}Mn_{2.4}Ge_{2}$} ferromagnet from polarized and non-polarized neutron powder diffraction aided by density-functional theory calculations},
journal = {Journal of Magnetism and Magnetic Materials},
volume = {529},
pages = {167827},
year = {2021},
issn = {0304-8853},
doi = {https://doi.org/10.1016/j.jmmm.2021.167827},
url = {https://www.sciencedirect.com/science/article/pii/S0304885321001037},
author = {Zachary P. Tener and Vincent Yannello and Jenifer Willis and V. {Ovidiu Garlea} and Michael Shatruk}
}

@article{laves1942einige,
  title={{\"U}ber einige neue Vertreter des NiAs-Typs und ihre kristallchemische Bedeutung},
  author={Laves, F and Wallbaum, HJ},
  journal={Zeitschrift f{\"u}r angewandte Mineralogie},
  volume={4},
  pages={17--46},
  year={1942}
}

@article{yasukochi1961magnetic,
author = {Yasuk\={o}chi ,K\={o} and Kanematsu ,Kazuo and Ohoyama ,Tetuo},
title = {{Magnetic Properties of Intermetallic Compounds in Iron-Germanium System} : {$\rm Fe_{1.67}Ge$} and {$\rm FeGe_2$}},
journal = {Journal of the Physical Society of Japan},
volume = {16},
number = {3},
pages = {429-433},
year = {1961},
doi = {10.1143/JPSJ.16.429},
URL = {https://doi.org/10.1143/JPSJ.16.429},
eprint = {https://doi.org/10.1143/JPSJ.16.429}
}

@book{gopal2012specific,
  title={Specific heats at low temperatures},
  author={Gopal, Erode},
  year={2012},
  publisher={Springer Science \& Business Media}
}

@article{Arrot_1957,
  title = {Criterion for Ferromagnetism from Observations of Magnetic Isotherms},
  author = {Arrott, Anthony},
  journal = {Phys. Rev.},
  volume = {108},
  issue = {6},
  pages = {1394--1396},
  numpages = {0},
  year = {1957},
  month = {Dec},
  publisher = {American Physical Society},
  doi = {10.1103/PhysRev.108.1394},
  url = {https://link.aps.org/doi/10.1103/PhysRev.108.1394}
}

@article{BANERJEE196416,
title = {On a generalised approach to first and second order magnetic transitions},
journal = {Physics Letters},
volume = {12},
number = {1},
pages = {16-17},
year = {1964},
issn = {0031-9163},
doi = {https://doi.org/10.1016/0031-9163(64)91158-8},
url = {https://www.sciencedirect.com/science/article/pii/0031916364911588},
author = {B.K. Banerjee}
}

@article{Fisher_1967,
doi = {10.1088/0034-4885/30/2/306},
url = {https://dx.doi.org/10.1088/0034-4885/30/2/306},
year = {1967},
month = {jul},
publisher = {},
volume = {30},
number = {2},
pages = {615},
author = {M E Fisher},
title = {The theory of equilibrium critical phenomena},
journal = {Reports on Progress in Physics}
}

@book{stanley1971phase,
  title={Phase transitions and critical phenomena},
  author={Stanley, H Eugene},
  volume={7},
  year={1971},
  publisher={Clarendon Press, Oxford}
}

@article{KAUL19855,
title = {Static critical phenomena in ferromagnets with quenched disorder},
journal = {Journal of Magnetism and Magnetic Materials},
volume = {53},
number = {1},
pages = {5-53},
year = {1985},
issn = {0304-8853},
doi = {https://doi.org/10.1016/0304-8853(85)90128-3},
url = {https://www.sciencedirect.com/science/article/pii/0304885385901283},
author = {S.N. Kaul}
}

@article{Fisher_Nickel_1972,
  title = {Critical Exponents for Long-Range Interactions},
  author = {Fisher, Michael E. and Ma, Shang-keng and Nickel, B. G.},
  journal = {Phys. Rev. Lett.},
  volume = {29},
  issue = {14},
  pages = {917--920},
  numpages = {0},
  year = {1972},
  month = {Oct},
  publisher = {American Physical Society},
  doi = {10.1103/PhysRevLett.29.917},
  url = {https://link.aps.org/doi/10.1103/PhysRevLett.29.917}
}

@article{Widom1964,
    author = {Widom, B.},
    title = {Degree of the Critical Isotherm},
    journal = {The Journal of Chemical Physics},
    volume = {41},
    number = {6},
    pages = {1633-1634},
    year = {1964},
    month = {09},
    issn = {0021-9606},
    doi = {10.1063/1.1726135},
    url = {https://doi.org/10.1063/1.1726135}
}

@article{Fischer_kaul2002,
  title = {Critical magnetic properties of disordered polycrystalline ${\mathrm{Cr}}_{75}{\mathrm{Fe}}_{25}$ and ${\mathrm{Cr}}_{70}{\mathrm{Fe}}_{30}$ alloys},
  author = {Fischer, S. F. and Kaul, S. N. and Kronm\"uller, H.},
  journal = {Phys. Rev. B},
  volume = {65},
  issue = {6},
  pages = {064443},
  numpages = {12},
  year = {2002},
  month = {Jan},
  publisher = {American Physical Society},
  doi = {10.1103/PhysRevB.65.064443},
  url = {https://link.aps.org/doi/10.1103/PhysRevB.65.064443}
}

@article{kanematsu1963,
author = {Kanematsu ,Kazuo and Yasuk\={o}chi ,K\={o} and Ohoyama ,Tetuo},
title = {Magnetic Properties of {(Fe, Co)1.67Ge} and {(Fe, Ni)1.67Ge}},
journal = {Journal of the Physical Society of Japan},
volume = {18},
number = {10},
pages = {1429-1436},
year = {1963},
doi = {10.1143/JPSJ.18.1429},
URL = {https://doi.org/10.1143/JPSJ.18.1429},
eprint = {https://doi.org/10.1143/JPSJ.18.1429}
}

@article{Fisher1974,
  title = {The renormalization group in the theory of critical behavior},
  author = {Fisher, Michael E.},
  journal = {Rev. Mod. Phys.},
  volume = {46},
  issue = {4},
  pages = {597--616},
  numpages = {0},
  year = {1974},
  month = {Oct},
  publisher = {American Physical Society},
  doi = {10.1103/RevModPhys.46.597},
  url = {https://link.aps.org/doi/10.1103/RevModPhys.46.597}
}

@article{Ohta2009,
  title = {Anomalous magnetization in the layered itinerant ferromagnet {LaCoAsO}},
  author = {Ohta, Hiroto and Yoshimura, Kazuyoshi},
  journal = {Phys. Rev. B},
  volume = {79},
  issue = {18},
  pages = {184407},
  numpages = {5},
  year = {2009},
  month = {May},
  publisher = {American Physical Society},
  doi = {10.1103/PhysRevB.79.184407},
  url = {https://link.aps.org/doi/10.1103/PhysRevB.79.184407}
}

@article{Shimizu1990,
author = {Shimizu ,Kazuaki and Maruyama ,Hiroshi and Yamazaki ,Hitoshi and Watanabe ,Hideki},
title = {Effect of Spin Fluctuations on Magnetic Properties and Thermal Expansion in Pseudobinary System {$\rm Fe_xCo_{1-x}Si$}},
journal = {Journal of the Physical Society of Japan},
volume = {59},
number = {1},
pages = {305-318},
year = {1990},
doi = {10.1143/JPSJ.59.305},
URL = {https://doi.org/10.1143/JPSJ.59.305},
eprint = {https://doi.org/10.1143/JPSJ.59.305}
}

@article{KHALANIYA2019118,
title = {When two is enough: On the origin of diverse crystal structures and physical properties in the {Fe-Ge} system},
journal = {Journal of Solid State Chemistry},
volume = {270},
pages = {118-128},
year = {2019},
issn = {0022-4596},
doi = {https://doi.org/10.1016/j.jssc.2018.10.030},
url = {https://www.sciencedirect.com/science/article/pii/S0022459618304638},
author = {Roman A. Khalaniya and Andrei V. Shevelkov}
}

@article{Lees1999,
  title = {Specific heat of {${\rm Pr_{0.6}(Ca_{1-\it x}Sr_{\it x})_{0.4}MnO_3}$ \,$(0\le \it x \le 1)$}},
  author = {Lees, M. R. and Petrenko, O. A. and Balakrishnan, G. and McK. Paul, D.},
  journal = {Phys. Rev. B},
  volume = {59},
  issue = {2},
  pages = {1298--1303},
  numpages = {0},
  year = {1999},
  month = {Jan},
  publisher = {American Physical Society},
  doi = {10.1103/PhysRevB.59.1298},
  url = {https://link.aps.org/doi/10.1103/PhysRevB.59.1298}
}

@article{KITAGAWA2020121188,
title = {Competition between ferromagnetic and antiferromagnetic states in {$\rm Al_{8.5−\it x}Fe_{23}Ge_{12.5 +\it x}$}  $(0 \leq x \leq 3)$},
journal = {Journal of Solid State Chemistry},
volume = {284},
pages = {121188},
year = {2020},
issn = {0022-4596},
doi = {https://doi.org/10.1016/j.jssc.2020.121188},
url = {https://www.sciencedirect.com/science/article/pii/S0022459620300189},
author = {Jiro Kitagawa and Genta Yakabe and Akinori Nakayama and Terukazu Nishizaki and Masami Tsubota}
}

@article{khalaniya2021magnetic,
  title={Magnetic structures of {$\rm Fe_{32+\delta}Ge_{33}As_{2}$} and {$\rm Fe_{32+\delta} Ge_{35-x}P_{x}$} intermetallic compounds: a neutron diffraction and $^{57}${Fe} {M{\"o}ssbauer} spectroscopy study},
  author={Khalaniya, R A and Sobolev, A V and Verchenko, V Yu and Tsirlin, A A and Senyshyn, A and Damay, F and Presniakov, I A and Shevelkov, A V},
  journal={Dalton transactions},
  volume={50},
  number={6},
  pages={2210--2220},
  year={2021},
  publisher={Royal Society of Chemistry},
  doi = {https://doi.org/10.1039/D0DT03923C}
}

@article{Deiseroth2006,
author = {Deiseroth, Hans-Jörg and Aleksandrov, Krasimir and Reiner, Christof and Kienle, Lorenz and Kremer, Reinhard K.},
title = {{$\rm Fe_3GeTe_2$} and {$\rm Ni_3GeTe_2$} – {Two New Layered Transition-Metal Compounds: Crystal Structures, HRTEM Investigations, and Magnetic and Electrical Properties}},
journal = {European Journal of Inorganic Chemistry},
volume = {2006},
number = {8},
pages = {1561-1567},
doi = {https://doi.org/10.1002/ejic.200501020},
url = {https://chemistry-europe.onlinelibrary.wiley.com/doi/abs/10.1002/ejic.200501020},
year = {2006}
}

@article{Zhang2020fe5xget2,
  title = {Itinerant ferromagnetism in van der Waals {$\rm Fe_{5-\it x}GeTe_{2}$}  crystals above room temperature},
  author = {Zhang, Hongrui and Chen, Rui and Zhai, Kun and Chen, Xiang and Caretta, Lucas and Huang, Xiaoxi and Chopdekar, Rajesh V. and Cao, Jinhua and Sun, Jirong and Yao, Jie and Birgeneau, Robert and Ramesh, Ramamoorthy},
  journal = {Phys. Rev. B},
  volume = {102},
  issue = {6},
  pages = {064417},
  numpages = {6},
  year = {2020},
  month = {Aug},
  publisher = {American Physical Society},
  doi = {10.1103/PhysRevB.102.064417},
  url = {https://link.aps.org/doi/10.1103/PhysRevB.102.064417}
}

@article{May2016fe3xgete2,
  title = {Magnetic structure and phase stability of the van der Waals bonded ferromagnet ${\mathrm{Fe}}_{3\ensuremath{-}x}{\mathrm{GeTe}}_{2}$},
  author = {May, Andrew F. and Calder, Stuart and Cantoni, Claudia and Cao, Huibo and McGuire, Michael A.},
  journal = {Phys. Rev. B},
  volume = {93},
  issue = {1},
  pages = {014411},
  numpages = {11},
  year = {2016},
  month = {Jan},
  publisher = {American Physical Society},
  doi = {10.1103/PhysRevB.93.014411},
  url = {https://link.aps.org/doi/10.1103/PhysRevB.93.014411}
}

@article{Das1994kondo,
  title = {{Kondo-lattice behavior of the interstitial alloys ${\mathrm{CeFe}}_{\mathit{x}}{\mathrm{Ge}}_{2}$}},
  author = {Das, I. and Sampathkumaran, E. V. and Hirota, K. and Ishikawa, M.},
  journal = {Phys. Rev. B},
  volume = {49},
  issue = {5},
  pages = {3586--3588},
  numpages = {0},
  year = {1994},
  month = {Feb},
  publisher = {American Physical Society},
  doi = {10.1103/PhysRevB.49.3586},
  url = {https://link.aps.org/doi/10.1103/PhysRevB.49.3586}
}

@misc{etde_5753394,
title = {Crystal structure of ternary germanides {$R\rm Fe_6Ge_6$} (\textit{R} -- {Sc, Ti, Zr, Hf, Nd}) and {$R\rm Co_6Ge_6$} (\textit{R}= {Ti, Zr, Hf})},
author = {Olenich, R R and Akselrud, L G and Yarmolyuk, Yu R},
volume = {2},
journal = {AC},
place = {Ukraine},
year = {1981},
month = {Feb}
}

@article{SCHOBINGERPAPAMANTELLOS199859,
title = {The {Fe} ordering in {RFe6Ge6} compounds with non-magnetic {R (R=Y, Lu, Hf)} studied by neutron diffraction and magnetic measurements},
journal = {Journal of Alloys and Compounds},
volume = {267},
number = {1},
pages = {59-65},
year = {1998},
issn = {0925-8388},
doi = {https://doi.org/10.1016/S0925-8388(97)00548-3},
url = {https://www.sciencedirect.com/science/article/pii/S0925838897005483},
author = {P Schobinger-Papamantellos and K.H.J Buschow and F.R {de Boer} and C Ritter and O Isnard and F Fauth}
}

@article{MAZET201379,
title = {Magnetic properties of {MgFe6Ge6}},
journal = {Solid State Communications},
volume = {159},
pages = {79-83},
year = {2013},
issn = {0038-1098},
doi = {https://doi.org/10.1016/j.ssc.2013.01.027},
url = {https://www.sciencedirect.com/science/article/pii/S0038109813000604},
author = {T. Mazet and V. Ban and R. Sibille and S. Capelli and B. Malaman}
}

@article{VENTURINI199299,
title = {Crystallographic data and magnetic properties of {RT6Ge6} compounds {(R= Sc, Y, Nd, Sm, Gd, Lu; T= Mn, Fe)}},
journal = {Journal of Alloys and Compounds},
volume = {185},
number = {1},
pages = {99-107},
year = {1992},
issn = {0925-8388},
doi = {https://doi.org/10.1016/0925-8388(92)90558-Q},
url = {https://www.sciencedirect.com/science/article/pii/092583889290558Q},
author = {G. Venturini and R. Welter and B. Malaman}
}

@article{Karna2021scfege,
  title = {Helical magnetic order and Fermi surface nesting in noncentrosymmetric {ScFeGe}},
  author = {Karna, Sunil K. and Tristant, D. and Hebert, J. K. and Cao, G. and Chapai, R. and Phelan, W. A. and Zhang, Q. and Wu, Y. and Dhital, C. and Li, Y. and Cao, H. B. and Tian, W. and Dela Cruz, C. R. and Aczel, A. A. and Zaharko, O. and Khasanov, A. and McGuire, M. A. and Roy, A. and Xie, W. and Browne, D. A. and Vekhter, I. and Meunier, V. and Shelton, W. A. and Adams, P. W. and Sprunger, P. T. and Young, D. P. and Jin, R. and DiTusa, J. F.},
  journal = {Phys. Rev. B},
  volume = {103},
  issue = {1},
  pages = {014443},
  numpages = {15},
  year = {2021},
  month = {Jan},
  publisher = {American Physical Society},
  doi = {10.1103/PhysRevB.103.014443},
  url = {https://link.aps.org/doi/10.1103/PhysRevB.103.014443}
}

@Article{D5DT00654F,
author ="Khalaniya, Roman A. and Verchenko, Valeriy Yu. and Mironov, Andrei V. and Samarin, Alexander N. and Bogach, Alexey V. and Kulchu, Aleksandr N. and Polevik, Alexey O. and Wei, Zheng and Dikarev, Evgeny V. and Stern, Raivo and Shevelkov, Andrei V.",
title  ="Spin reorientation and magnetic frustration in {$\rm Fe_{32+\delta}Ge_{35−\it x}Si_{\it x}$} with a kagome lattice broken by crystallographic intergrowth",
journal  ="Dalton Trans.",
year  ="2025",
volume  ="54",
issue  ="20",
pages  ="8317-8330",
publisher  ="The Royal Society of Chemistry",
doi  ="10.1039/D5DT00654F",
url  ="http://dx.doi.org/10.1039/D5DT00654F"}

\end{document}


\author{P.L.S. Cambalame\corref{cor1}\fnref{label1,label2} \orcidlink{0000-0002-6959-8931}}
\ead{phinifolo@fis.uc.pt}
\author{B.J.C. Vieira \orcidlink{0000-0002-6536-9875}\fnref{label3}}
\author{J.C. Waerenborgh\orcidlink{0000-0001-6171-4099}\fnref{label3}}
\author{P.S.P. da Silva \orcidlink{0000-0002-6760-4517}\fnref{label1}}
\author{J.A. Paix\~{a}o \orcidlink{0000-0003-4634-7395}\corref{cor2}\fnref{label1}}
\ead{jap@fis.uc.pt}
\cortext[cor2]{}
\affiliation[label1]{organization={CFisUC, Department of Physics, University of Coimbra},
		   city={Coimbra},
		   postcode={3004-516},
		   country={Portugal}}
\affiliation[label2]{organization={Department of Physics, Eduardo Mondlane University}, 
		   city={Maputo}, 
		   postcode={0101-11}, 
		   country={Mozambique}}
\affiliation[label3]{organization={Centro de Ci\^{e}ncias e Tecnologias Nucleares, DECN, Instituto Superior T\'{e}cnico, Universidade de Lisboa},
		   city={Bobadela},
		   postcode={2695-066},
		   state={LRS},
		   country={Portugal}}

	\title{Supplementary Information. $\rm CrFe_2Ge_2$: Investigation of novel ferromagnetic material of $\rm Fe_{13}Ge_{8}$-type crystal structure} 
	
\maketitle
\section{Second synthesis route}
A second sample, used for the remaining magnetic and transport property measurements was synthesized via a modified route (sample $\# 2$). The reactants were annealed at 600 $^{\circ}$C for 48 hours, then heated up to 750 $^{\circ}$C, after 12 hours, the temperature was raised up to 1150 $^{\circ}$C and maintained for 8 hours, followed by the cooling from 1150 $^{\circ}$C to 950 $^{\circ}$C at rate of 0.1 $^{\circ}$C/min.

\begin{figure}[!htpb]
	\captionsetup{font=small}
		\centering
		\includegraphics[width =.6\linewidth]{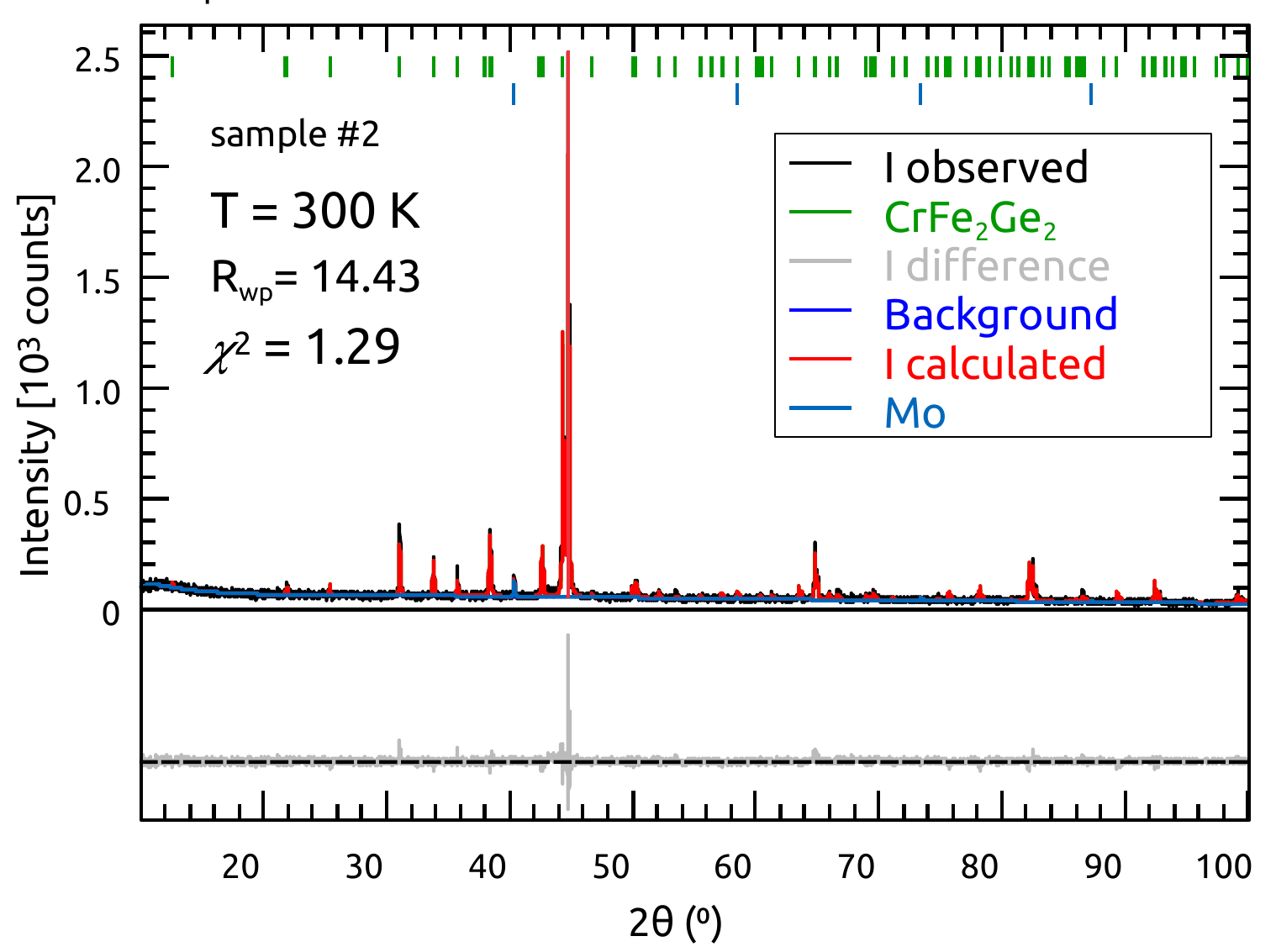}
	\caption{Powder X-ray diffraction spectra of $\rm CrFe_2Ge_2$ second sample and the results of Profex Rietvield refinement.  Tick marks indicate X-ray diffraction peaks of $\rm CrFe_2Ge_2$ and $\rm Mo$.} 
	\label{fig:pxrd}
\end{figure}
Powder X-ray diffractogram analysis reveals a minor $\rm Mo ~(< 1\%)$ impurity peak centered 40.40 $^\circ$ and the remaining spectra corresponds to the target phase, $\rm CrFe_{2}Ge_{2}$. The identified impurity exhibited negligible influence on the properties of the target material.

\section{Magnetic characterization of sample $\#1$ and $\#2$}

Sample $\#1$ was used for M\"{o}ssbauer spectroscopy, single-crystal X-ray diffraction and initial magnetic characterization. Sample $\#2$ was used for detailed magnetic  characterization, transport and specific heat measurements.

\begin{figure*}[!htpb]
	\captionsetup{font=small}
	\centering
	\begin{subfigure}{.45\linewidth}
		\centering
		\includegraphics[width = .88\linewidth]{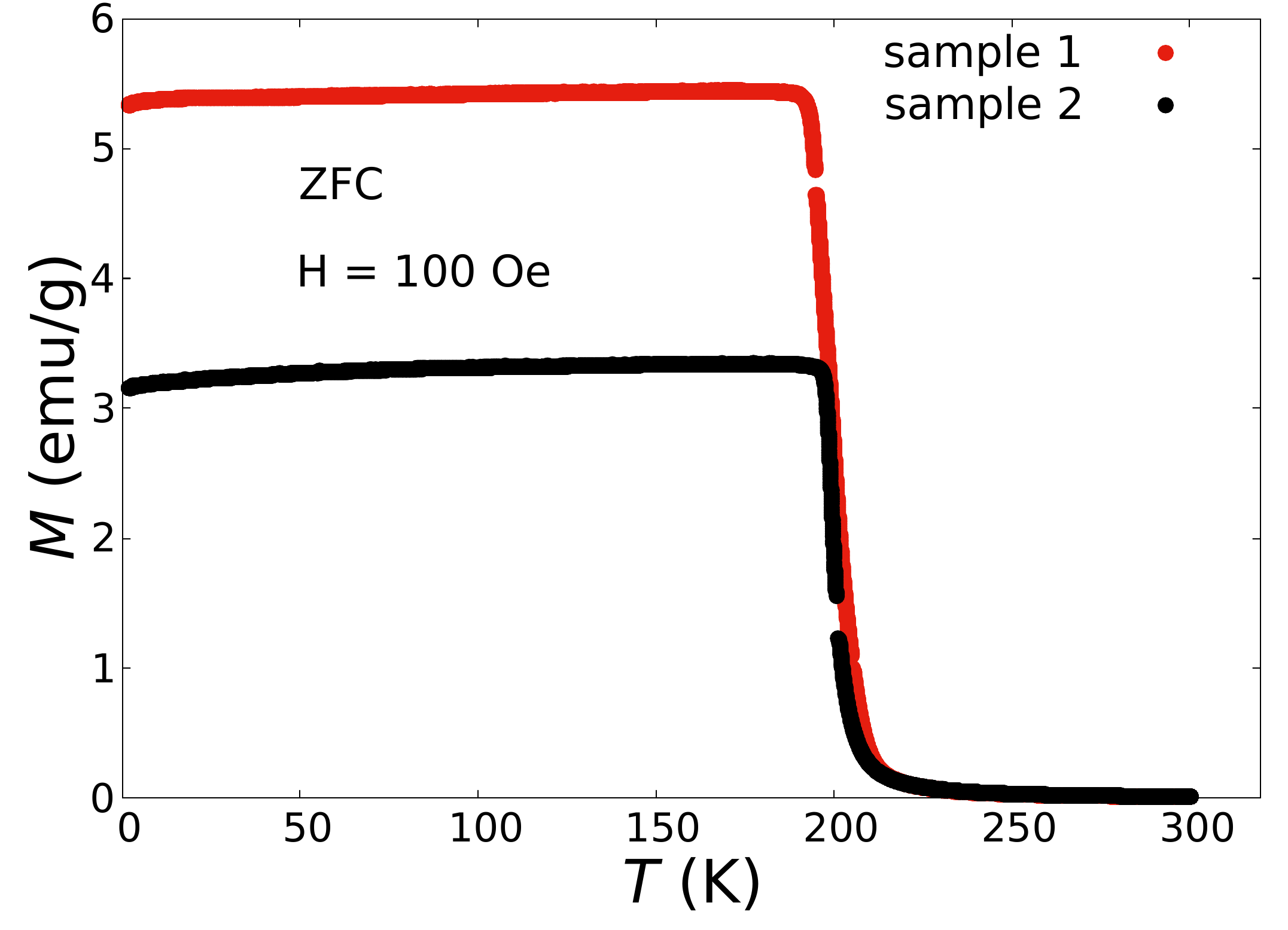}
		\caption{}
	\end{subfigure}%
	\vspace{0.41em}
	\begin{subfigure}{.45\linewidth}
		\centering
		\includegraphics[width =0.88\linewidth]{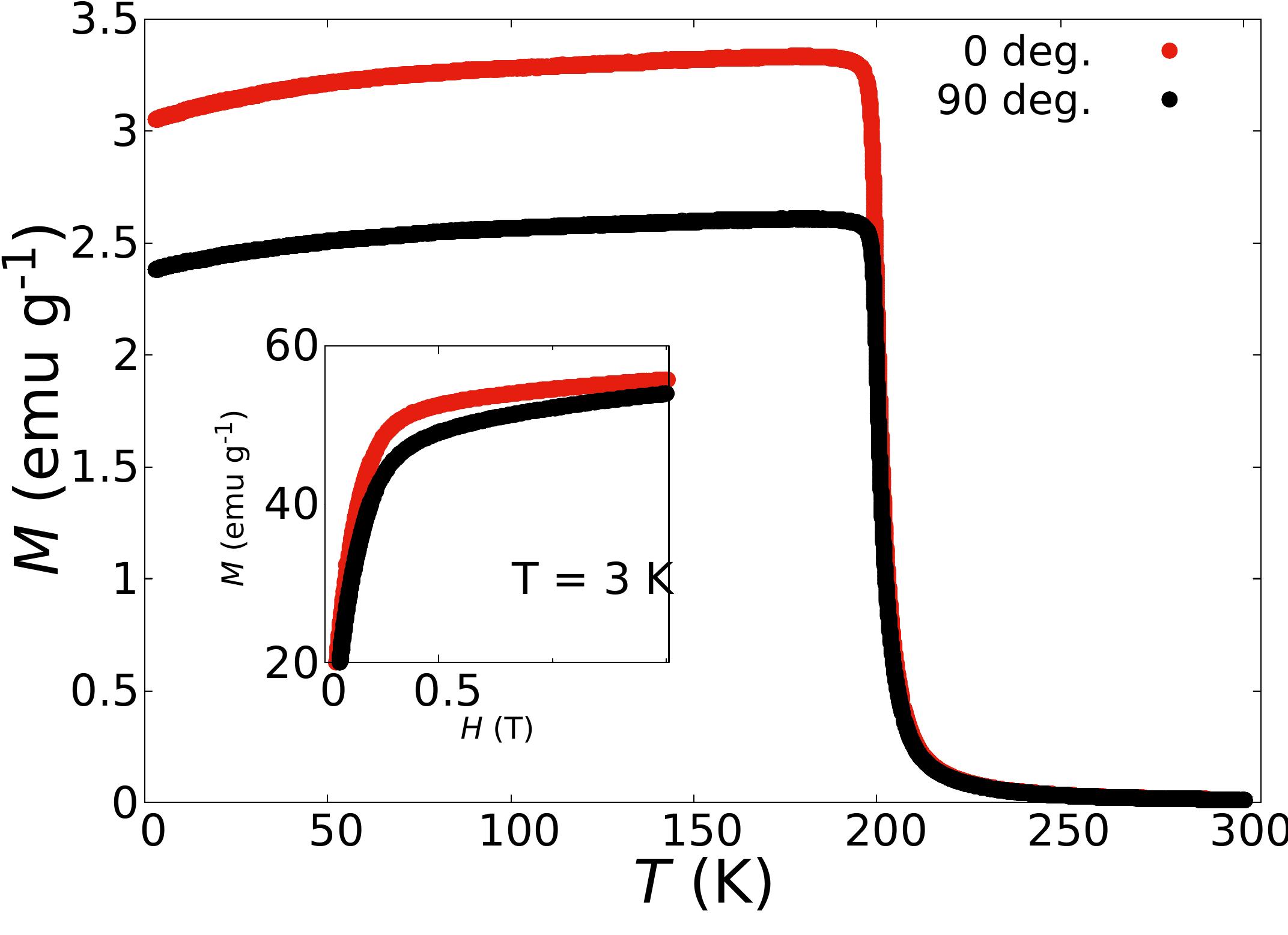}
		\caption{}
	\end{subfigure}%
	
	\hspace{1em}
	\begin{subfigure}{.45\linewidth}
		\centering
		\includegraphics[width =0.88 \linewidth]{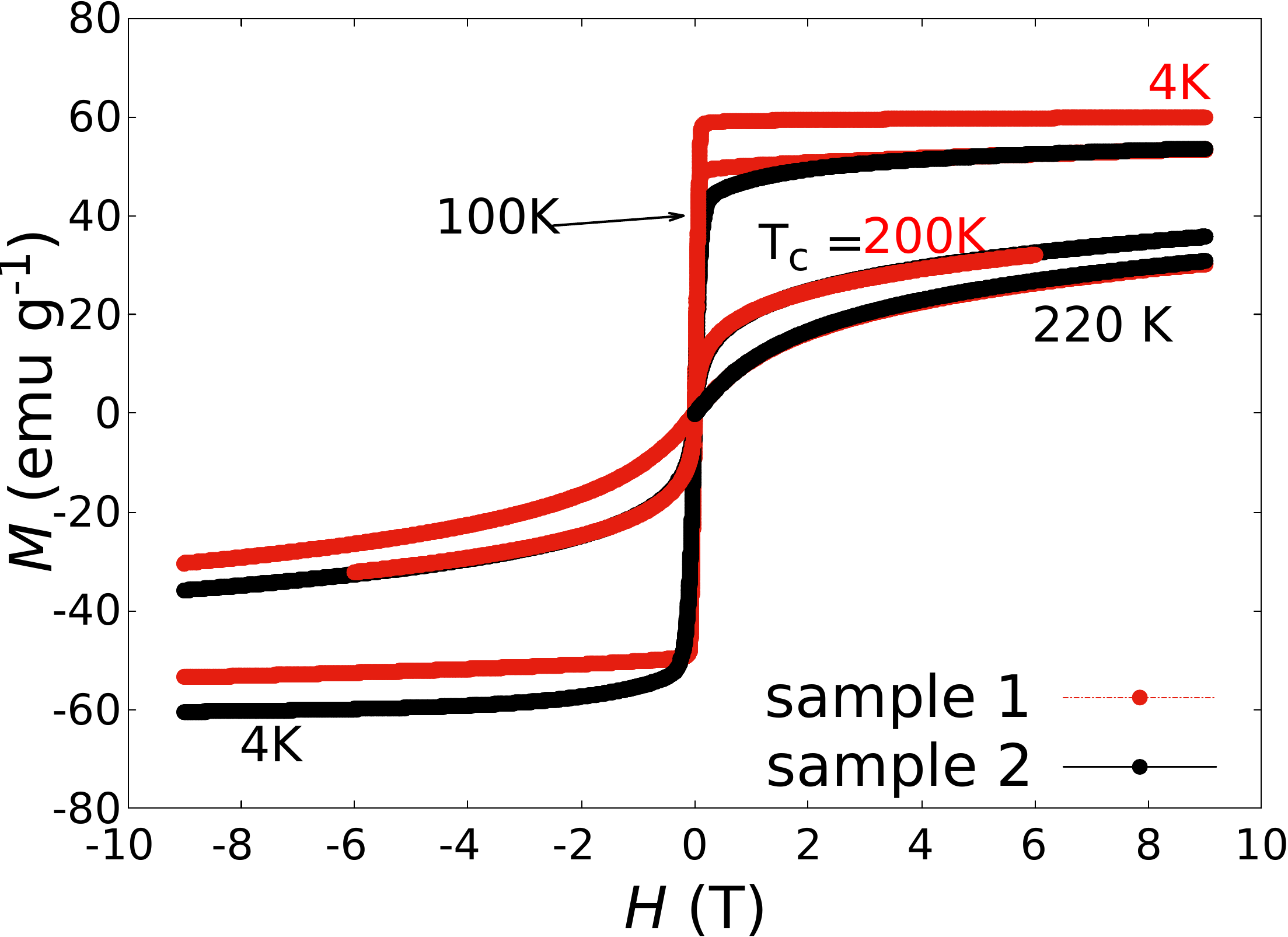}
		\caption{}
	\end{subfigure}%
	\vspace{0.41em}
	\begin{subfigure}{.45\linewidth}
		\centering
		\includegraphics[width = 0.88\linewidth]{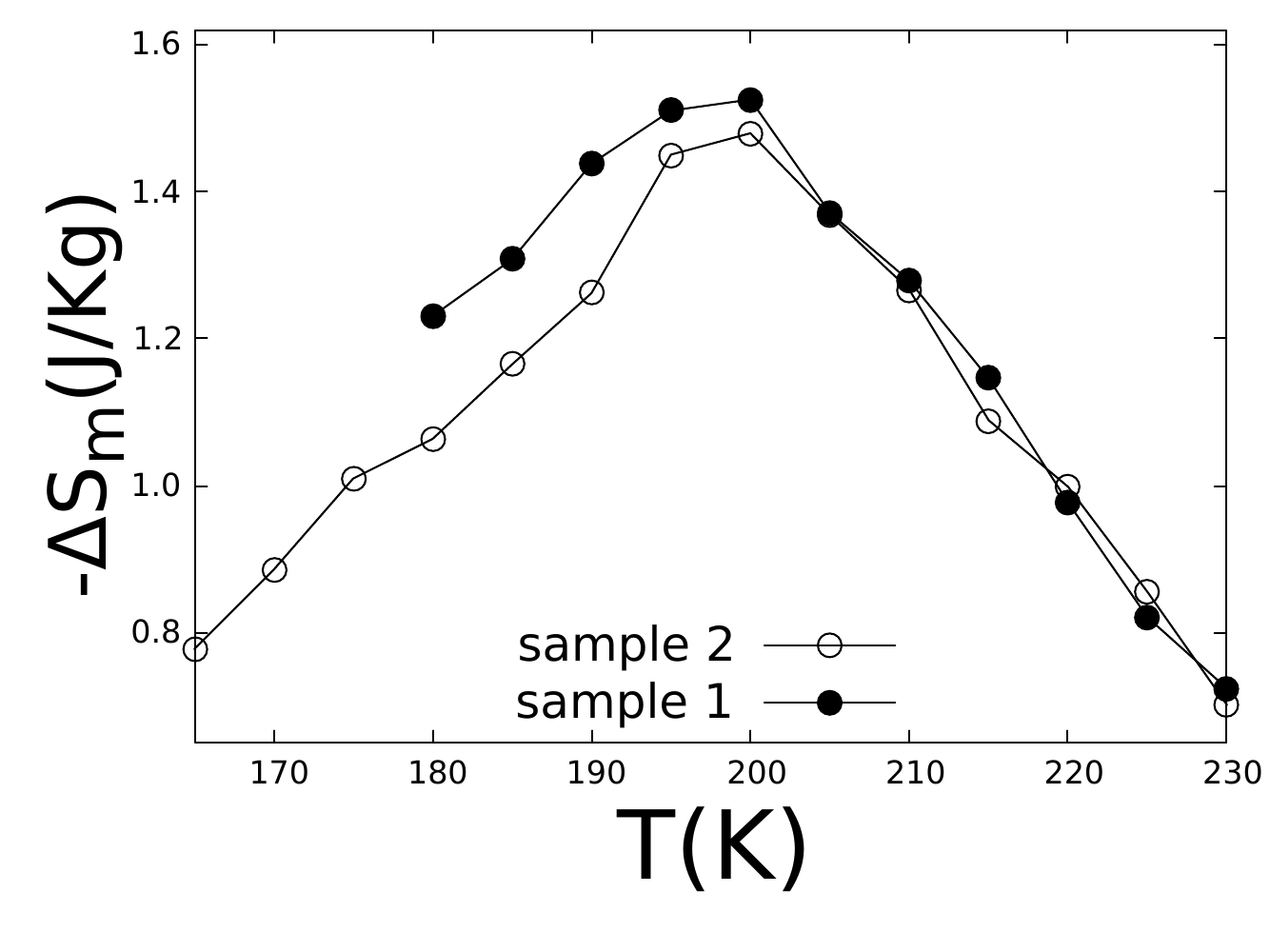}
		\caption{}
	\end{subfigure}%
	\caption{(a) ZFC Magnetization as function of temperature of sample 1 and 2 of $\rm CrFe_2Ge_2$. (b) sample 2: ZFC $M(T)$ for $H = 100 \rm Oe$ applied field and $M(H)$ isotherm  measured for 0 and 90 degree rotated sample. (c) $M(H)$ isotherms at various temperatures at magnetic field \textit{H} of up to 9 T for sample 1 and 2. (d) Magnetic entropy for samples 1 and 2 of $\rm CrFe_2Ge_2$.\label{fig:mag_data_s1}}
\end{figure*}

The matching and consistency of magnetic properties (Figure \ref{fig:mag_data_s1}) together with X-ray studies confirm that  both samples correspond  to the same material and that the contribution of $\rm Mo$ to the properties of second sample is negligible. The difference in magnetization intensity (Figure \ref{fig:mag_data_s1} (a)) can be attributed to the better crystallization and to measurements on sample $\#1$ being performed closer to the easy axis. This anisotropy is also evident when the thermomagnetic curves of $\rm CrFe_2Ge_2$ are recorded after 90$^{\circ}$ rotation from the initial orientation, as shown in Figure \ref{fig:mag_data_s1} (b).
The sharper curvature of the $M(H)$ isotherms for sample $\#1$ compared to sample $\#2$ (Figure \ref{fig:mag_data_s1} (c)) can likewise be explained by the average easy-axis orientation and superior crystallization of sample $\#1$. These differences become more pronounced deeper in the magnetically ordered region, whereas in the paramagnetic and critical regions the magnetic isotherms of the two samples overlap. These distinctions also account for the differences in full-width at half-maximum of $-\Delta S_m(T)$ (Figure \ref{fig:mag_data_s1} (d)).

\section{Critical behavior}
\begin{figure*}[!htpb]
	\captionsetup{font=small}
	\centering
	\begin{subfigure}{.5\linewidth}
		\centering
		\includegraphics[width = .98\linewidth]{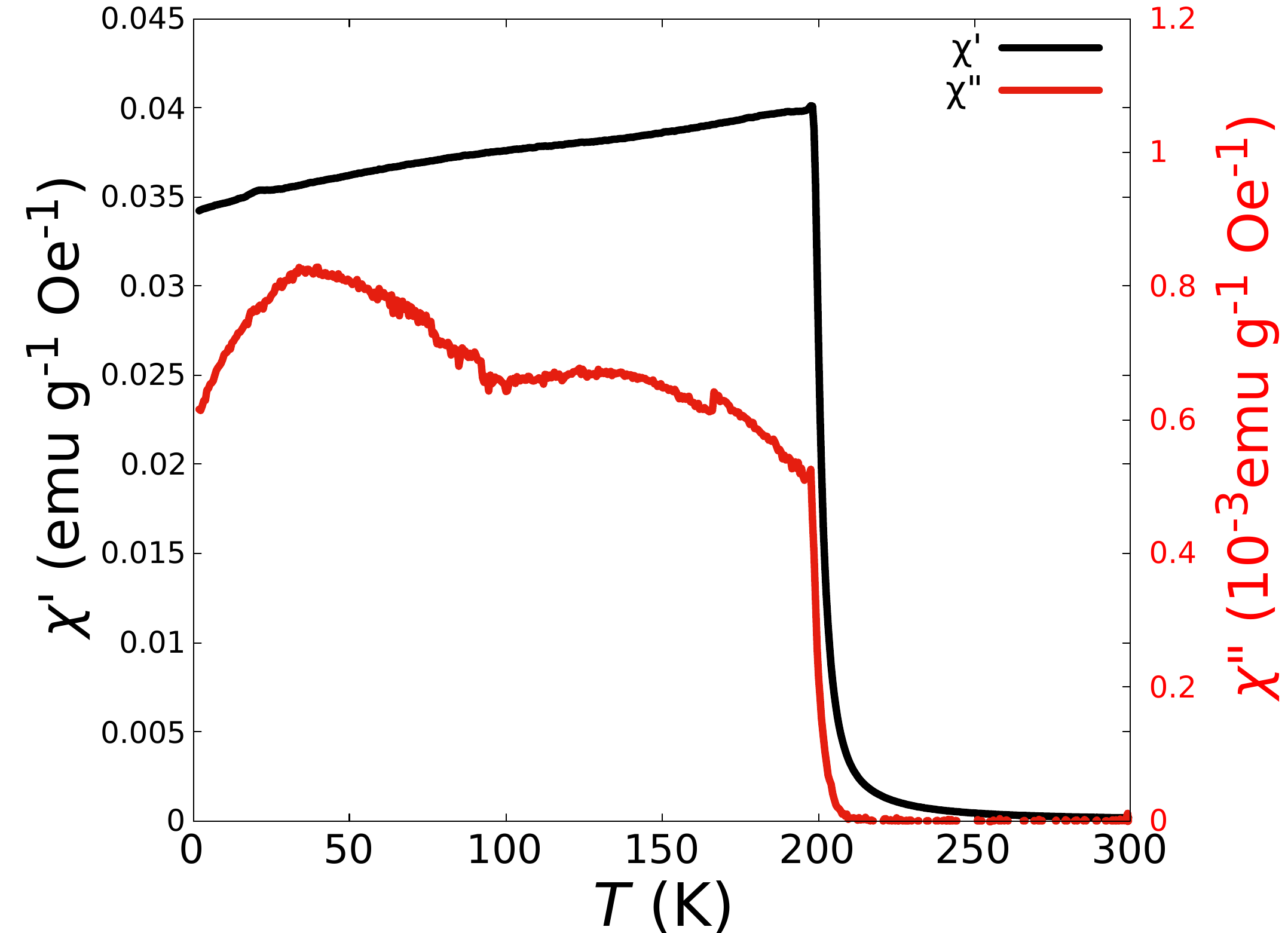}
		\caption{}
	\end{subfigure}%
	\vspace{0.41em}
	\begin{subfigure}{.5\linewidth}
		\centering
		\includegraphics[width =0.98\linewidth]{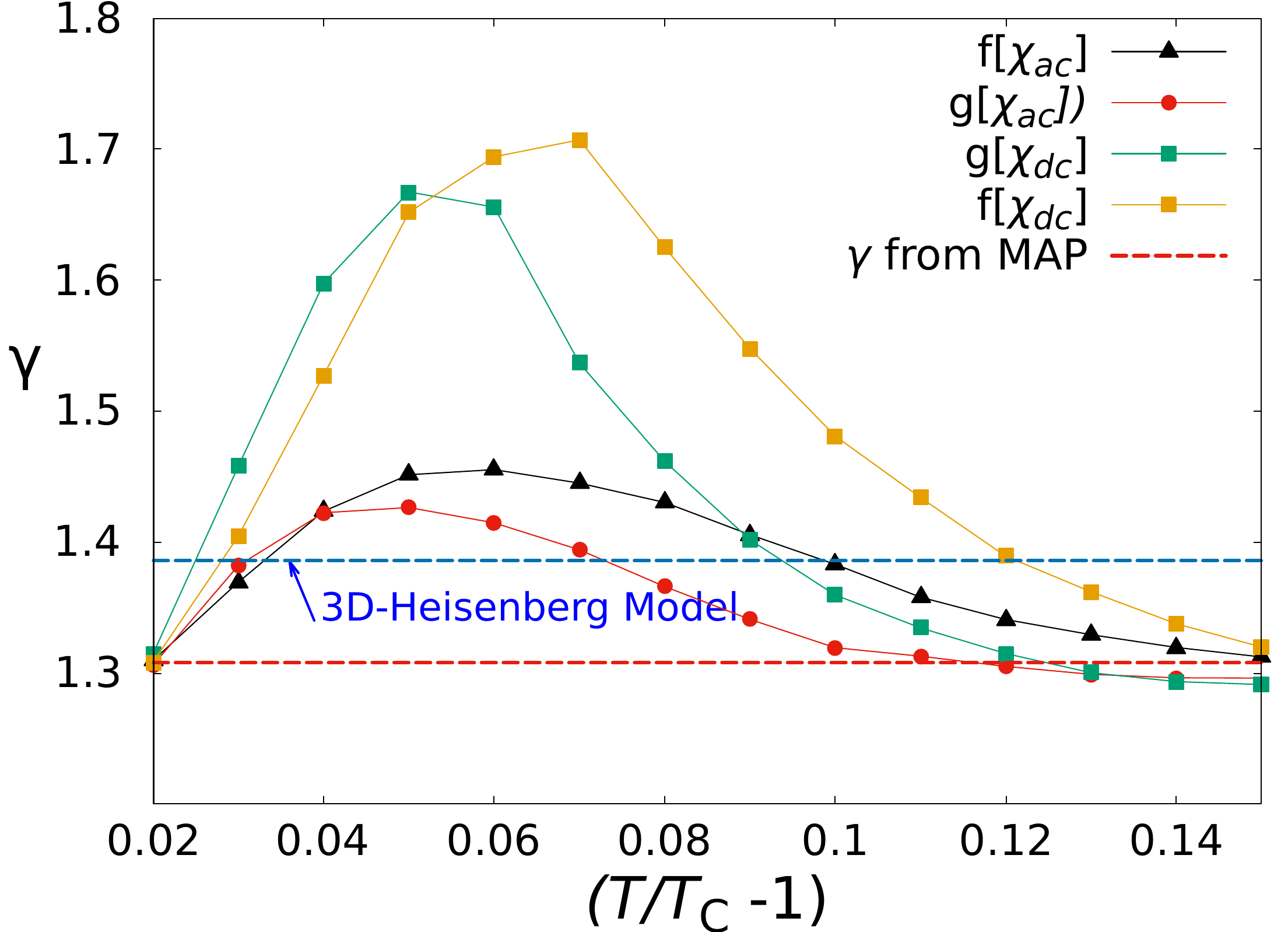}
		\caption{}
\end{subfigure}%
\caption{(a) Real part $\chi'$ and imaginary part $\chi''$ of $\rm CrFe_2Ge2$ magnetic susceptibility. Measurements parametes: $f = 500~ \rm Hz$, AC driving field of $\rm 5 ~Oe$. (b) $\gamma$ as function of range of fit using the equations (\ref{eq_si:f_func}) and (\ref{eq_si:g_func}) on  ac measurements data $(f[\chi_{ac}] ~\mathrm{and} ~g[\chi_{ac}])$, and on dc measurement data with  $\rm 50 ~Oe$, $(f[\chi_{dc}] ~\mathrm{and} ~g[\chi_{dc}])$ .\label{fig:mag_data_s3}}
\end{figure*}

The fitting of inverse initial susceptibility was performed using the paramagnetic region of ac susceptibility mesured at a frequency of 500 Hz with an ac driving field of 5 Oe. The real $\chi '$ and imaginary $\chi''$ parts are displayed in the Figure \ref{fig:mag_data_s3} (a). To obtain $\gamma$, we first applied a single power-law fit of the form $\chi_0^{-1} = x_0\epsilon^{\gamma}$. However, this approach render inconsistent results as $\epsilon$ varied and yielded poor fits near $T_{\rm C}$. To address this issue, we included the leading nonanalytic correction to scaling [1].
\begin{equation}
	\chi_0^{-1} = x_0 \epsilon ^{\gamma}(1 + x_1 \epsilon ^{\Delta})^{-1} ~~, \epsilon > 0,
\label{eq_si:gamma}
\end{equation}

Considering a narrow range of fit $(\epsilon < 0.1)$, and  a constant $\Delta$ ($\Delta = 0.55$ for 3D-Heisenberg systems [2]) and the condition $|x_1 \epsilon ^{\Delta}| < 1$, the expression can be expanded  as convergent  series of $n+1$ terms:

\begin{equation}
	q=-x_1\epsilon^{\Delta}
\end{equation}

\begin{equation}
	\frac{1}{1-q}=\sum_{0}^{\infty}q^n= 1 -x_1\epsilon^{\Delta} + x_1^2\epsilon^{2\Delta} - x_1^3\epsilon^{3\Delta} + ...
\label{eq_si:convergence}
\end{equation}
From this series, we update the equation \ref{eq_si:gamma} to

\begin{equation}
		\chi_0^{-1} = x_0 \epsilon ^{\gamma} - x_0 \epsilon ^{\gamma}(x_1\epsilon^{\Delta} - x_1^2\epsilon^{2\Delta} + x_1^3\epsilon^{3\Delta} + ...)	
\label{eq_si:gamma3}
\end{equation}

Two versions of expression (\ref{eq_si:gamma3}) were implemented, first:
\begin{equation}
	f(T) =x_0 \epsilon ^{\gamma} - x_0 \epsilon ^{\gamma}(x_1\epsilon^{\Delta} - x_1^2\epsilon^{2\Delta} + x_1^3\epsilon^{3\Delta})~~,\epsilon = (T/T_C - 1) 
\label{eq_si:f_func}
\end{equation}

and a simplified version in which the second and higher order terms are reduced to a single constant $G_1 =  x_0 \epsilon ^{\gamma}(x_1\epsilon^{\Delta} - x_1^2\epsilon^{2\Delta} + x_1^3\epsilon^{3\Delta} + ...)$, under the assumption that the convergence condition of equation (\ref{eq_si:convergence}) remains valid. This yields

\begin{equation}
	g(T) = x_0 \epsilon ^{\gamma} - G_1
\label{eq_si:g_func}
\end{equation}

Within the asymptotic critical region of $\epsilon \leq 0.02$, the values of $\gamma$ obtained from all implemented functions converges to $\gamma_{eff} \approx 1.31$ within the error limits. Furthermore, the quality of the fit does not improve when the equation \ref{eq_si:f_func} is used instead of the equation \ref{eq_si:g_func}.

\section*{References}
[1] M. Seeger, S. N. Kaul, H. Kronmüller, R. Reisser, \textit{Asymptotic critical behavior of Ni}, Phys. Rev. B 51 (1995) 12585–12594.

[2] C. Bagnuls, C. Bervillier, \textit{Nonasymptotic critical behavior from field theory at d = 3: The disordered-phase case}, Phys. Rev. B 32 (1985) 7209–
7231.